\documentclass[14pt]{article}
\pdfoutput=1
\usepackage{amsmath,amssymb,amsfonts,amsthm,bm,bbm,cancel,wasysym}
\usepackage{epsfig,graphics,graphicx,epstopdf}
\usepackage{array,booktabs,colortbl,colordvi,multirow}
\usepackage{colordvi,color,xcolor}
\usepackage{hyperref}
\usepackage{rotating}

\usepackage{verbatim}
\usepackage{cite}
\usepackage{subfig}
\usepackage{setspace}
\usepackage{url}
\usepackage[percent]{overpic}
\usepackage{slashed}
\usepackage{authblk}
\usepackage{xspace}
\usepackage{fullpage}
\usepackage{hyperref}

\def\ie{{\it i.e.}}
\def\eg{{\it e.g.}}
\def\etc{{\it etc}}

\def\to{\rightarrow}

\newskip\zatskip \zatskip=0pt plus0pt minus0pt
\def\matth{\mathsurround=0pt}
\def\lsim{\mathrel{\mathpalette\atversim<}}
\def\gsim{\mathrel{\mathpalette\atversim>}}
\def\atversim#1#2{\lower0.7ex\vbox{\baselineskip\zatskip\lineskip\zatskip
  \lineskiplimit 0pt\ialign{$\matth#1\hfil##\hfil$\crcr#2\crcr\sim\crcr}}}

\begin{document}


\begin{flushright}
SLAC-PUB-17728\\
\today
\end{flushright}
\vspace*{5mm}

\renewcommand{\thefootnote}{\fnsymbol{footnote}}
\setcounter{footnote}{1}

\begin{center}

{\Large {\bf Towards UV-Models of Kinetic Mixing and Portal Matter IV: Quartification 
}}\\

\vspace*{0.75cm}

{\bf Thomas G. Rizzo}~\footnote{rizzo@slac.stanford.edu}

\vspace{0.5cm}

{SLAC National Accelerator Laboratory}\\ 
{2575 Sand Hill Rd., Menlo Park, CA, 94025 USA}

\end{center}
\vspace{.5cm}


\begin{abstract}
\noindent  

As is well-known, Trinification, \ie, the extension of the Standard Model (SM) to $[SU(3)]^3=SU(3)_c\times SU(3)_L\times SU(3)_R$ as occurs in $E_6$ models, allows for a partial unification 
of the gauge forces even though quarks and leptons remain in separate multiplets so that no heavy gauge or scalar fields exist which can generate proton decay.  The extension of this idea 
to Quartification, by including an additional $SU(3)'$ factor, has also been considered in the literature maintaining the basic attributes of Trinification but now allowing, \eg, for a more symmetric 
treatment of quarks and leptons at the price of new matter fields and gauge interactions.  In this paper, we will consider this $SU(3)'$ to be the `dark' gauge group, now containing the familiar 
$U(1)_D$ subgroup, under which the SM fields are all neutral, which is associated with kinetic mixing (KM) and the existence of a light, $\lsim 1 $ GeV dark photon. This setup naturally predicts the 
existence of color-singlet portal matter (PM) fields, carrying both electromagnetic and $U(1)_D$ dark charges, that are necessary to generate this KM at the 1-loop level and whose masses are 
directly tied with those of the many new gauge bosons that originate from the extended gauge sector. In this paper, after a discussion of the detailed structure of this model, we present a broad 
survey of the collider phenomenology of the large set of new fields that must necessarily arise from this setup in a simplified version involving only a single generation of fermions. We demonstrate 
that several new signatures may be anticipated at the LHC as well as at future hadron and lepton colliders if such models are realized in nature.
\end{abstract}

\vspace{0.5cm}
\renewcommand{\thefootnote}{\arabic{footnote}}
\setcounter{footnote}{0}
\thispagestyle{empty}
\vfill
\newpage
\setcounter{page}{1}



\section{Introduction and Background Discussion}

The Standard Model (SM) of particle physics, though very successful, faces a number of significant obstacles in explaining the world that we see. One of the most challenging and long standing 
of these is the nature of dark matter (DM) for which there is no SM candidate: what is it and how does it interact with us, if at all, beyond the obvious gravitational interactions by which we know 
it to exist?  A perhaps reasonable expectation is that DM and SM fields will interact rather weakly through some new, as yet unknown, mediator which itself is not a part of the SM and by whose 
action the DM obtains the the value of the relic density as measured by Planck\cite{Planck:2018vyg}. It is not unreasonable to ask how the SM might be extended to account for this possibility 
and how a more unified description which includes these new interactions may be achieved. 

Such questions in one form or another have been under discussion for quite some time and the consideration of various DM candidates now extends back several decades. As is well-known,
the searches for these `traditional' DM candidates, such as the QCD axion\cite{Kawasaki:2013ae,Graham:2015ouw,Irastorza:2018dyq} and the family of weakly interacting massive 
particles, \ie, WIMPS\cite{Arcadi:2017kky,Roszkowski:2017nbc},  continue to push downwards into ever lower cross section regimes and wider over larger ranges of possible masses with 
increasing sensitivities. Unfortunately, these direct detection experiments, as well as those searching for indirect signatures or via the direct production of DM at the 13 TeV 
LHC\cite{LHC,Aprile:2018dbl,Fermi-LAT:2016uux,Amole:2019fdf,LZ:2022ufs}, have all so far produced negative results, thus excluding increasing large regions of the parameter spaces of 
many specific models. These results and others have, over the last decade or so, led the community to greatly expand upon this set of traditional candidates with many new ideas for the possible 
nature of DM and its interactions with the SM. During this time interval it has become clear that both the coupling strength of DM to (at least some of) the fields of the SM as well as its possible 
mass can both span previously unexpected large ranges\cite{Alexander:2016aln,Battaglieri:2017aum,Bertone:2018krk,Cooley:2022ufh,Boveia:2022syt,Schuster:2021mlr} which will require a wide 
variety of very broad and very deep searches to provide even partially adequate coverage.  In a similar fashion, it has been found that many distinct types of interactions between DM and 
the fields of the SM are possible, many of which are best classified by the use of the effective field theory approach employing `portals'. Such portals can produce interaction structures which are 
either renormalizable (\ie, dimension $\leq 4$) or non-renormalizable (\ie, dimension $> 4$) depending upon the specific setup. For these models to work, an additional set of new fields also 
need to be introduced which act as mediators linking the SM to the DM and potentially also to an enlarged, non-trivial, Dark Sector of which the DM candidate is only the lightest stable state - 
possibly due to, \eg, the existence of some new, at least approximately conserved, quantum number.

Among the various classes of these portals, significant attention has been given in the literature to the renormalizable kinetic mixing (KM)/vector portal\cite{KM,vectorportal,Gherghetta:2019coi} 
scenario which is based upon the existence of a new dark gauge interaction and which has significant model building flexibility.  In such a scenario, one finds that even in the most simple 
realization, over significant ranges of the model parameters, it is possible for the DM to reach its observed abundance via the usual WIMP-like thermal freeze-out 
mechanism\cite{Steigman:2015hda,Saikawa:2020swg}. Unlike in traditional WIMP models, however, this will now occur only for sub-GeV DM masses by employing this previously mentioned 
new non-SM dark gauge interaction that, because of its weakness, has so far evaded detection by other means. The simplest construction of this kinds has only a few moving pieces
assuming only the existence of this new gauge interaction based on the $U(1)_D$ gauge group, with a coupling $g_D$, under which it it is postulated that the SM fields are all neutral, 
implying that they do not carry dark charges, \ie, thus $Q_D=0$. The new $U(1)_D$ gauge boson is then termed as the `dark photon' (DP) \cite{Fabbrichesi:2020wbt,Graham:2021ggy}, which 
we will generally denote as $D$. In the usual setup, in order to obtain the observed relic density by thermal means, this new $U(1)_D$ is assumed to be spontaneously broken at or 
below the scale of $\sim$ few GeV scale and thus both the DM and DP will have comparable masses. This symmetry breaking is most simply accomplished via the (sub-)GeV scale vev(s) of 
at least one (if not several) new scalar(s), the dark Higgs(es), similar to the symmetry breaking which occurs in the SM. While the DP couples to DM it does not do so at tree-level with SM 
fields but, within this framework, is generated via renormalizable kinetic mixing (KM) at the 1-loop level between the $U(1)_D$ and the SM $U(1)_Y$ gauge fields and whose strength is 
then described by a small, dimensionless parameter, $\epsilon$. Since the SM fields have $Q_D=0$ and DM has no SM charges, these loops must arise from a set of new fields, 
usually being vector-like (VL) fermions and/or complex scalars, which we call Portal Matter (PM) \cite{Rizzo:2018vlb,Rueter:2019wdf,Kim:2019oyh, Rueter:2020qhf,Wojcik:2020wgm,Rizzo:2021lob,Rizzo:2022qan,Wojcik:2022rtk,Rizzo:2022jti,Rizzo:2022lpm,Wojcik:2022woa,Carvunis:2022yur,Verma:2022nyd,Rizzo:2023qbj,Wojcik:2023ggt}, 
that must carry {\it both} both SM and $U(1)_D$ dark charges. Transforming back to the familiar canonically normalized gauge fields to remove the KM and after both the SM and $U(1)_D$ gauge 
symmetries are broken, this now leads to an effective coupling of the DP to SM fields of the form $\simeq e\epsilon Q_{em}$ (up to small correction terms or order $m_D^2/m_Z^2<<1$).  
Further, for both DM and a DP being sub-GeV in mass, one finds that the magnitude of the parameter $\epsilon$ is constrained by experiment to very roughly lie in the 
$\epsilon \sim 10^{-(3-4)}$ range, a number which we might have already expected from it originating due to a loop. Importantly, in such a setup, for $p-$wave annihilating DM or for pseudo-Dirac 
DM with a sufficiently large mass splitting, it is found the rather tight constraints arising from the CMB on DM annihilation into electromagnetically interacting final states can also be 
avoided\cite{Planck:2018vyg,Slatyer:2015jla,Liu:2016cnk,Leane:2018kjk} for an overlapping range of model parameters.

Finally, another interesting feature of this class of setups, in the conventionally chosen normalization and in the IR limit, is that the parameter $\epsilon$ can be determined in terms of the properties 
of the PM fields that appear in these vacuum polarization-like graphs and is given by the sum
\begin{equation}
\epsilon =\frac{g_D  e}{24\pi^2} \sum_i ~\big(\eta_i   N_{c_i}Q_{em,i}Q_{D_i}\big)~ ln \frac{m^2_i}{\mu^2}\,,
\end{equation}
where $e(g_D)$ is the usual QED ($U(1)_D$) gauge coupling and $m_i(Q_{em,i},Q_{D_i}, N_{c_i})$ are the mass (electric charge, dark charge, number of colors) of the $i^{th}$ PM field.  
We note that, \eg, $\eta_i=1(1/2)$ if the PM particle is assumed to be a chiral fermion (complex scalar){\footnote{In the model framework below, there are two gauge bosons that can also play 
this PM role and so will also contribute to the sum above as we will discuss below.}}. 
We then see that if the condition 
\begin{equation}
\sum_i ~\eta_i   N_{c_i}Q_{em,i}Q_{D_i}=0\,,
\end{equation}
is also satisfied, as might perhaps be expected within a fully unified description, then $\epsilon$ also becomes a finite and, in principle, a calculable quantity. Such an observation may 
already whet our appetite to search for such a more enveloping framework for the KM scenario. However, as we will see in the discussion below, this condition is not automatically satisfied in 
this only partially unified model without the introduction of some additional (likely scalar) fields beyond this minimal setup.

It seems to be advantageous to go beyond this effective theory to further our understanding of how this (apparently) simple KM mechanism fits together into a single scheme with the 
SM, something that we, and others, have begun to examine in pathfinder mode employing various bottom-up and top-down approaches in a recent series of papers 
\cite{Rizzo:2018vlb,Rueter:2019wdf,Rueter:2020qhf,Wojcik:2020wgm,Rizzo:2021lob,Rizzo:2022qan,Wojcik:2022rtk,Rizzo:2022jti,Rizzo:2022lpm,Wojcik:2022woa,Rizzo:2023qbj,Wojcik:2023ggt}. 
Two common features that one finds from following our general approach are the embedding of the $U(1)_D$ dark abelian symmetry into some larger, non-abelian $G_{Dark}$, 
\eg, with an $SU(2)_I\times U(1)_{Y_I}$\cite{Rueter:2019wdf} gauge symmetry\cite{Bauer:2022nwt} being its simplest manifestation and the appearance of at least some of the SM fields 
in common gauge group representations with the PM fields. In such setups, the PM mass generation is generally the result of the $G_{Dark} \to U(1)_D$ symmetry breaking and 
so, with $O(1)$ Yukawa couplings, will share a similar overall mass scale with the associated heavy gauge bosons.  This was seen explicitly in Ref.\cite{Rueter:2019wdf} whose PM content and the 
$G_{Dark}=SU(2)_I\times U(1)_{Y_I}$ gauge group were both inspired by $E_6$\cite{e6,Hewett:1988xc}. More recently, we extended this idea of a more general product group 
$G=G_{SM}\times G_{Dark}$ from a top-down perspective\cite{Rizzo:2022lpm} and also via an augmentation of $G_{SM}$ to that of the Left-Right Symmetric Model 
(LRM)\cite{Pati:1974yy,Mohapatra:1974hk,Mohapatra:1974gc,Senjanovic:1975rk,book} while maintaining $G_{Dark}=SU(2)_I\times U(1)_{Y_I}$ wherein the PM and RH-neutrino mass 
scales were shown to be related. Interestingly, one generally finds that attempts to `unify' the SM with the KM portal naturally brings in some of the others portals as well, \eg, the Higgs and neutrino 
portals will also frequently appear. 

Following the cue from these previous studies, in this paper we consider identifying $G_{SM}=SU(3)_c\times SU(3)_L \times SU(3)_R$, \ie, the Trinification\cite{trinif} subgroup of $E_6$, 
while also assuming that $G_{Dark}=SU(3)'$, thus forming a Quartification model. Such a class of partially unified models has been discussed in the literature for both DM as well as in 
other contexts\cite{quart} and in some cases allows for greater flexibility 
than many conventional unification approaches involving a single gauge group or a product of two groups. While the general setup that we will obtain following this path will share some easily 
recognizable common characteristics with one or both of the previously examined $G_{Dark}=SU(2)_I\times U(1)_{Y_I}$ model classes, it will present us with numerous new and interesting features   
that we will begin to explore in the present work. Unfortunately, due to the necessarily numerous new fields of all varieties that will appear in the current study (even ignoring the fact of there being 
three generations) it is quite difficult to make many specific phenomenological predictions that can provide more that suggestive tests for this setup beyond a few specific examples without 
making further assumptions 
about, \eg, the relative sizes/orderings of the large mass scale vevs.  These specific choices will induce different possible paths of symmetry breaking down to the SM$\times U(1)_D$ thus leading to 
quite different, quantitative predictions for, \eg, the masses and, in some cases, the couplings of the many new color singlet gauge bosons within this setup. However, there remains many 
aspects that most of these paths will share, at least at the semi-quantitative level, and we will concentrate our efforts in exploring those phenomenological tests here. Specifically, we will examine 
the capabilities of the LHC as well as future hadron and lepton colliders to explore the essential components of this setup: the new gauge bosons and their interplay with the VL fermions and 
SM fields.  As we will see, this 
scenario shares many of the features that we had earlier encountered in our analyses of one or both of the previously examined $E_6$ and Pati-Salam motivated models but with different (and 
sometimes simultaneous) aspects of additional simplicity and complexity. Unfortunately, while a step forward, this type of setup is not yet fully unified in the traditional sense.

The outline of this paper is as follows:  Following the present Introduction and Background discussion, in Section 2 we will present a broad outline of the model setup and framework, setting 
the stage for the analysis in the later Sections. Sections 3 and 4 will then individually examine the various fermionic sectors of this setup, \ie,  the generation of the Dirac and Majorana fermion 
masses together with the corresponding mixings between the PM and SM/LRM fermion fields. The KM and gauge symmetry breaking which takes place in several distinct steps at a hierarchy 
of mass scales and the resulting gauge boson masses and mixings that will be important at the electroweak scale and below will then be discussed separately for the non-hermitian and hermitian 
gauge fields in Sections 5 and 6, respectively. A further examination of a sample of some of the (mostly collider oriented) phenomenological implications and tests of this scenario that were 
not touched upon in any detail earlier as part of the model development will then be presented in Section 7. As we'll see below, given the numerous moving parts for this model, it is quite difficult to 
capture all of the interesting new physics potential of this setup within this introductory survey in more than a suggestive manner. Finally, a summary and discussion of our results, possible 
future avenues of exploration and our subsequent conclusions can then be found in Section 8.


\section{Overview of Model Setup and Framework}

Although the details of the fermion embedding, the interpretation and low energy phenomenology of the $SU(3)'$ gauge group will differ here from that in much of the literature, the 
essential features and the basic mechanics of the general Quartification model setup will remain unaltered\cite{quart}. However, in what follows we will {\it not}, \eg, assume that an additional $Z_4$ 
exchange symmetry among the four $SU(3)$ groups is also present and we will not be directly interested in specific unification issues at very high mass scales. This freedom, \eg, allows for some 
asymmetric treatment in the necessary symmetry breaking of the various gauge group factors as we will encounter in the discussion below.  In terms of the diagonal generators 
of the $SU(3)_c\times SU(3)_L\times SU(3)'\times SU(3)_R$ gauge group, the electric charge in this version of Quartification will be given by the somewhat uncommon, but highly symmetric 
decomposition 
\begin{equation}
Q_{em}= T_{3L}-\frac{Y_L}{2}+T_{3R}-\frac{Y_R}{2}+T_3'-\frac{Y'}{2}\,,
\end{equation}
where $Y_a/2=T_{8a}/\sqrt 3$ with $a=(L,R,')$. Similarly, as we will see below, the dark charge, $Q_D$, associated with the familiar $U(1)_D$ gauge group as described above, is just 
given by a linear combination of the two diagonal $SU(3)'$ generators{\footnote {The origin of this expression will be explained shortly.}}, up to an overall sign convention, as 
\begin{equation}
Q_D= T_3'-\sqrt 3 T_8'\,.
\end{equation}

As noted above, since we will not be concerned here with the potential vast wealth of flavor phenomenology associated with this kind of setup, we will limit ourselves in our discussion below to 
the consideration of a only single fermion generation so that we can emphasize the corresponding wealth of non-flavor physics. To this end, we will label all the SM fermion fields by their first 
generation labels. Interestingly, a respectable fraction of the phenomenology that we will encounter in the present model has already been seen in our earlier work on PM
\cite{Rizzo:2018vlb,Rueter:2019wdf,Rueter:2020qhf,Wojcik:2020wgm,Rizzo:2021lob,Rizzo:2022qan,Wojcik:2022rtk,Rizzo:2022jti,Rizzo:2022lpm,Wojcik:2022woa,Rizzo:2023qbj,Wojcik:2023ggt} 
but in other and generally more simplified settings. Although related at the `unification' scale, at accessible collider energies we will treat the 3 gauge couplings for $3_L3'3_R$, \ie, $g_L,g',g_R$ 
as generally independent quantities although, for ease and clarity of presentation at some later points, we will assume they have a common value.

In terms of this $SU(3)_c\times SU(3)_L\times SU(3)'\times SU(3)_R$ Quartification gauge group, which hereafter we will frequently abbreviate simply as $3_c3_L3'3_R$, helped by the left-right  
gauge symmetries and the VL nature of the new fields, the 36, two-component, left-handed fermion fields of a single generation are found to transform in an anomaly-free manner with equal numbers 
of triplets and anti-triplets for each of the $SU(3)$ group factors combined with singlets. In addition to the 15 SM fields plus the familiar RH-neutrino, there will also now be 8 new VL fermions: 
one electroweak singlet, VL quark (\ie, 6 degrees of freedom) plus 3 charged and 4 neutral VL leptons (\ie, 14 degrees of freedom). This full set of fermions for a single generation is given by the 
sum of the four representations: 
\begin{equation}
[q]~(3,\bar 3,1,1)+[l]~(1,3,\bar 3,1)+[l^c]~(1,1,3,\bar 3)+[q^c]~(\bar 3,1,1,3)\,,
\end{equation}
where we want to especially note that all of the color (anti-)triplet fields are seen to be singlets under the new $SU(3)'$ group so that in this framework the PM fields must necessarily be color 
singlets, \eg, VL leptons and/or complex scalars. The $q,q^c$ fields, being $SU(3)'$ singlets,  are the familiar, canonical ones that also appear in $E_6$-inspired Trinification models\cite{trinif}:
\begin{equation}
[q]=\begin{pmatrix} d & u & h\\ d & u & h\\ d& u & h\\ \end{pmatrix},~~~~[q^c]=\begin{pmatrix} d^c & d^c & d^c\\ u^c & u^c & u^c\\ h^c& h^c & h^c\\ \end{pmatrix}\,,
\end{equation} 
with $h$ being, to those familiar with $E_6$ phenomenology\cite{e6,Hewett:1988xc}, a $Q_{em}=-1/3$, color-triplet, electroweak iso-singlet. However, unlike in other previously considered 
PM scenarios, here $h$ is ${\it not}$ 
a PM field as it (and all the quarks) is a $Q_D=0$ singlet under the $SU(3)'$ group and so is simply a `conventional' VL quark as has been sought at the LHC, excluding a 
mass below $\sim 1.2$ TeV assuming decays\cite{Alves:2023ufm} only to the third generation SM fermions{\footnote {One possible difference is that in this setup $h$ may be sufficiently massive 
as to have decays through non-SM mediators or into non-SM final states but these are unlikely to be dominant.}}. Clearly, the fact that quarks do not carry any $SU(3)'$ charges will greatly impact 
the production of the various new states we will encounter below at hadron colliders.  The $l,l^c$ fields, on the other hand, are here somewhat different due to, 
amongst other things, the relationship between the diagonal $SU(3)'$ generators and $Q_{em}$ as well as the necessity of keeping SM fields free of any $Q_D$ charge. Below, for 
these `leptonic' fields, and following somewhat closely the nomenclature of, \eg,  the recent work of Ma in Ref.\cite{quart} to make contact with the existing literature, we see that :  
\begin{equation}
[l]=\begin{pmatrix} \nu & E_1^c & N_1\\ e & N_1^c & E_1\\ S_1& E_3^c & S_2\\ \end{pmatrix},~~~~[l^c]=\begin{pmatrix} \nu^c & e^c & S_1^c\\ E_2 & N_2 & E_3\\ N_2^c& E_2^c & S_2^c\\ \end{pmatrix}\,,
\end{equation} 
where the $S_{1,2}(E_3)$ and their conjugates have $|Q_{em}|=0(1)$ and are $2_L2_R$ singlets, $(N_1,E_1)[(N_2,E_2)]^T$ with both helicities together forming VL doublets with 
$Q_{em}=0,-1$, respectively, under $2_L[2_R]$ and are singlets under $2_R[2_L]$, while $(\nu,e)^T+{\rm h.c.}$ are just the conventional SM fields with the addition of the RH-neutrino. Since 
the usual $U(1)_D$ gauge group 
must be a diagonal subgroup of the $SU(3)'$ group, the familiar dark charge $Q_D$ must be a linear combination of the two diagonal generators, $T_{3,8}'$, chosen in a way such that 
the SM fields $\nu, e$, by construction, have $Q_D=0$. Normalizing to unit charges, this uniquely fixes, up to a sign choice, the relationship $Q_D= T_3'-\sqrt 3 T_8'$ as given above when 
these generators act 
on an $SU(3)'$ (anti-)triplet. From this we see that $Q_D(S_1)=0$ whereas all the other new lepton fields, $N_i,E_i,S_2$, will all carry $Q_D=-1$ and those which also have $Q_{em}\neq 0$ can  
be identified as true PM fields.  Generally, lepton-like PM will, of course, be somewhat more difficult to produce and observe at a hadron collider as previously noted.

Lastly, although we will not be directly discussing unification issues in the analysis that follows, we note in passing that with definition of $Q_{em}$ given above, as well as the details of 
the fermion representation structure, we can calculate the value of $\sin^2 \theta_w$ at the scale at which the product of the gauge symmetries, $G=[SU(3)]^4$, begin to break via the relation
\begin{equation}
\sin^2 \theta_w=\frac{Tr~ T_{3L}^2}{Tr~Q_{em}^2}=\frac{1}{4}\,,
\end{equation}
obtaining an interesting result that has been previously discussed in the Quartification literature\cite{quart} and which is quite suggestive. Similarly, one finds that the analog of 
this quantity for the dark gauge group,  $s_I^2=\sin^2 \theta_I$, as was employed in our earlier analyses of the $SU(2)_I \times U(I)_{Y_I}$, $E_6$-inspired model\cite{Rueter:2019wdf}, 
satisfies a similar relationship and is given by 
\begin{equation}
\sin^2 \theta_I=\frac{Tr~ T_3'^2}{Tr~Q_D^2}=\frac{1}{4}\,.
\end{equation}

Given the complexity of the setup that we will examine below, the opportunities for both KM and mass mixing between the DP and the many more massive neutral gauge boson states are 
rampant. Fortunately, the absence of bi-adjoint representations somewhat restricts this possibility. Of course, in the IR, this KM is in practice given simply by Eq.(1) above.

\section{Dirac Fermion Masses}

In order to generate the various fermion masses, we recall that we can form $SU(3)$ singlets either as a product of a ${\bf 3}$ with a ${\bf {\bar 3}}$ or by taking anti-symmetric products 
of three ${\bf 3}$'s or ${\bf {\bar 3}}$'s. This results in a somewhat restrictive, though sufficient, setup for the fermion mass terms and, as we will see, also the need to introduce an additional 
Higgs scalar representation, not coupling to the fermions, in order to properly generate all of the heavy gauge boson masses. In particular, as $SU(3)_c$ must remain unbroken, the 
independent set of Higgs fields that can be employed to generate the fermion masses can be at most three in number and must transform as bi-triplets under the remaining broken 
$SU(3)$ groups. Specifically, in order to obtain the Dirac masses for all of these fermions, it is sufficient to introduce the three Higgs fields, $H_i$, $i=1-3$, which transform in obvious manners 
with respect to the group $3_c3_L3'3_R$ based on the transformation properties of the fermion representations themselves as described above. A respectable fraction of these scalars will be 
eaten as they play the role of the Goldstone bosons for the many heavy gauge boson fields that we will encounter below. These $3\times 9=27$ complex scalars are given by
\begin{equation}
H_1=(1,3,1,\bar 3),~~~~H_2=(1,3,\bar 3,1),~~~~ H_3=(1,1,3,\bar 3)\,,
\end{equation}
so that the following Yukawa couplings (with repeated indices summed over) are allowed:
\begin{equation}
{\cal L}_y=y_1q_{ij}q^c_{jk}(H_1)_{ki}+y_2l_{ij}l^c_{jk}(H_1^\dagger)_{ki}+ \epsilon_{ijk}\epsilon_{\alpha \beta \gamma} \Big[y_3 l_{i\alpha}l_{j\beta}(H_2)_{k\gamma}+y_4 l^c_{i\alpha}l^c_{j\beta}
(H_3)_{k\gamma}\Big]+\rm{h.c.}\,.
\end{equation}
In the analysis presented here, for simplicity, we will ignore possible CP-violation and assume that all of these Yukawa couplings (as well as the many vevs that we will soon encounter below) are 
all real.  Note that {\it none} of the scalars appearing in representation $H_1$ carry a $Q_D\neq 0$ charge. More generally, knowing the relationships of both $Q_{em}$ and $Q_D$ to the relevant 
gauge group generators, we see that each of the $H_i$ is found to contain 5 neutral scalars so that, in total, it is possible to contemplate 15 distinct vevs associated with the various 
required mass scales. In particular, the following vevs for the neutral fields within these Higgs scalars, $H_i$, can be considered:
\begin{equation}
{\sqrt 2} <H_1>= \begin{pmatrix} v_1& 0& v_4\\ 0 & v_2 & 0\\ v_5& 0 & v_3\\ \end{pmatrix}\,,
\end{equation} 
which are solely responsible for quark masses and mixings but also contribute to the lepton masses as well, and  
\begin{equation}
\sqrt 2 <H_2>=\begin{pmatrix} u_1& 0 & x_1\\ 0 & x_2& 0\\ u_2& 0 & x_3\\ \end{pmatrix}~~~~\sqrt 2 <H_3>=\begin{pmatrix} u_3 & 0 & u_4\\ 0 & x_4 & 0\\ x_5& 0& x_6\\ \end{pmatrix}\,,
\end{equation} 
which govern the mass generation for the combined leptonic and PM sectors. 
All of the vevs, $v_i$, within $H_1$ correspond to scalars with $Q_D=0$ but, whereas $v_{1,2,4}$ will lead to the breaking of $SU(2)_L$, and so must be $\lsim 100$ GeV or so,  the vevs 
$v_{3,5}$ do not and either will simply lead to the breaking $SU(2)_R$ or to the more general breaking of $3_L3_R \to 2_L2_R1_{L+R}$ so we can easily imagine them being at the 
$\sim 10$ TeV mass scale or above. Similarly, while the vevs $u_{2-4}$ in $H_{2,3}$ all correspond with $Q_D=0$ scalars, they do not break $SU(2)_L$ and so are also expected to be large, again  
$\gsim 10$ TeV. On the other hand, the vev $u_1$, while still having $Q_D=0$, breaks $SU(2)_L$ and so, like $v_{1,2,4}$ above, is expected to be of order $\lsim 100$ GeV. The remaining 
six vevs in $H_{2,3}$, the $x_i$, all arise from scalars having $|Q_D|=1$ so will lead to the breaking of $U(1)_D$ and thus are required to lie at the much smaller $\lsim 1$ GeV scale.  Due to 
the hierarchy of scales of $\sim O(10^2)$ or possibly greater between the 3 sets of vevs, it will sometimes be convenient to treat their effects iteratively. Although rather obvious, it should be 
noted that whereas the vevs, 
$v_i$, will lead to a breaking of only $3_L3_R$, the vevs,  $u_i$, will instead break either $3_L3'$ or $3_R3'$ and so, in particular, link the scales associated with $SU(3)_R/SU(2)_R$ and 
$SU(3)'$ breaking, similar to those discussed in earlier work\cite{Rizzo:2023qbj}.  Table~\ref{vevtab} summarizes the quantum number and transformation properties of this large set of vevs that are 
responsible for, at least partly, both gauge symmetry breaking and fermion mass generation; those appearing in $H_1$ appear almost universally in the Quartification literature\cite{quart}.  Note 
that among the set of $v_i,u_i$ vevs, both the 
same and opposite signs for their values of $T_{3L},Y_L$ (and also $T_{3R},Y_R$) appear {\it but} only like sign values of $T_3',Y'$ appear; this is dictated by the requirement of $Q_D$ 
conservation. Once the $U(1)_D$ generator is broken by the small $x_i$ vevs we indeed see both the same as well as opposite signs of $T_3',Y'$ now appearing.  

Interestingly, we note that unless there exists some relationship among the various vevs it is difficult to break only one of the $SU(3)$ group factors without also breaking another unless some 
care is taken which can have a significant impact on the pattern of how all these symmetries must eventually be broken down to $U(1)_{em}$. Note that neglecting the $Q_D\neq 0$ vevs 
is an excellent first approximation in obtaining the fermion masses below.

\begin{table}[h]
\vspace*{0.5cm}
\caption{Higgs Scalar Vacuum Expectation Values}
\label{vevtab} 
The properties of the multiple vevs contained in the three Higgs scalar representations, $H_{1-3}$, that are discussed in the text and which are responsible for generating both Dirac and Majorana 
fermion masses and which contribute to gauge symmetry breaking. The horizontal line in the middle of the Table separates the $Q_D=0$ from the $Q_D \neq 0$ vevs acts as a guide to the eye. Here  
$Q_D= T_3'-\sqrt 3 T_8'$. It is to be noted that both of the vevs, $x_{2,4}$, will also violate lepton number by 2 units, \ie, are $\Delta L=2$, as is also discussed in the text.
\vspace*{0.25cm}
\begin{center}
\begin{tabular}{ l  c  c  c  c  c  c  c }
\hline
\hline		
vev & $T_{3L}$ & $Y_L/2$ & $T_{3R}$ & $Y_R/2$  & $T_3'$ & $Y'/2$ & $Q_D$  \\ 
\hline
$v_1$ & 1/2 & 1/6 & -1/2 & -1/6 & 0 & 0 & 0 \vspace{0.1cm}\\
$v_2$ & -1/2 & 1/6 & 1/2 & -1/6 & 0 & 0 & 0 \vspace{0.1cm}\\
$v_3$ & 0 & -1/3 &  0 & 1/3 & 0 & 0 & 0 \vspace{0.1cm}\\
$v_4$ & 1/2 & 1/6 & 0 & 1/3 & 0 & 0 & 0 \vspace{0.1cm}\\
$v_5$ & 0 & -1/3 & -1/2 & -1/6 & 0 & 0 & 0 \vspace{0.1cm}\\
$u_1$ & 1/2 & 1/6 & 0& 0 & -1/2 & -1/6 & 0 \vspace{0.1cm}\\
$u_2$ & 0 & -1/3 & 0 & 0 & -1/2 & -1/6 & 0 \vspace{0.1cm}\\
$u_3$ & 0 & 0 & -1/2 & -1/6 & 1/2 & 1/6 & 0 \vspace{0.1cm}\\
$u_4$ & 0 & 0 & 0 & 1/3 & 1/2 & 1/6 & 0 \vspace{0.1cm}\\
\hline
$x_1$ & 1/2 & 1/6 & 0 & 0 & 0 & 1/3 & -1 \vspace{0.1cm}\\
$x_2$ & -1/2 & 1/6 & 0 & 0 & 1/2 & -1/6 & 1 \vspace{0.1cm}\\
$x_3$ & 0 & -1/3 & 0 & 0 & 0 & 1/3 & -1 \vspace{0.1cm}\\
$x_4$ & 0 & 0 & 1/2 & -1/6 & -1/2 & 1/6 & 1 \vspace{0.1cm}\\
$x_5$ & 0& 0 & -1/2 & -1/6 & 0 & -1/3 & -1 \vspace{0.1cm}\\
$x_6$ & 0 & 0 & 0 & 1/3 & 0 & -1/3 & -1 \vspace{0.1cm}\\
\hline
\hline
\end{tabular}
\end{center}
\vspace*{0.5cm}
\end{table}

The mass terms generated by the $H_1$ vevs (all of which have $Q_D=0$) with a single Yukawa coupling for the quarks are the simplest to consider and take the form
\begin{equation}
\frac{y_1}{\sqrt 2}\Big( d^cdv_1+u^cuv_2+h^chv_3+h^cdv_4+d^chv_5\Big) +\rm{h.c.}\,,
\end{equation} 
where we see that, as might be expected, there is a single term for the $u$-type quarks whereas the $d,h$ fields undergo a mixing quite similar to what one finds in the $E_6$-inspired scenarios,  
forming a $2\times 2$ mass matrix of the form $D^cM_{-1/3}D$, where $D=(d,h)^T$ and 
\begin{equation}
M_{-1/3}=\frac{y_1}{\sqrt 2}~\begin{pmatrix} v_1 & v_5 \\ v_4 & v_3 \\ \end{pmatrix}\,,
\end{equation} 
that can be diagonalized as usual by a bi-unitary transformation, which in this very simple case can be expressed via the mixing angles between the two right-handed and two left-handed 
fields, respectively. To leading order in the small vev ratios, these are given by the expressions  

\begin{equation}
\tan 2\phi^{-1/3}_R \simeq \frac{2v_3v_5}{v_5^2-v_3^2}~~~~~~\tan 2\phi^{-1/3}_L \simeq \frac{-2(v_1v_5+v_4v_3)}{v_3^2+v_5^2}\,.
\end{equation} 
From these we see that we might expect $\phi^{-1/3}_L$ to be relatively small, $\lsim 10^{-2}$, due to the appearance of the ratio of the small to large vevs, whereas $\phi^{-1/3}_R$ may 
be significantly larger since only the large multi-TeV scale vevs appear to leading order in this particular case. The possible relative sizes of $v_{3,5}$ (as well as the other large vevs) will become 
further clarified when we discuss the gauge symmetry breaking and the corresponding various new gauge boson masses later below.

As is easily imagined, the mass terms for the leptonic sector are significantly more complex as they involve two fermion representations, can arise from all three sets of $Q_D=0$ Higgs vevs in the 
$<H_i>$ as expressed above, as well as involving multiple mass scales. We note that in the absence of any of the $Q_D\neq 0$ vevs we can uniques assign lepton number to all of the fields in 
both $l$ and $l^c$ and also to the corresponding scalars lying in $H_{2,3}$, \eg, it is easy to see that in this approximation all of the $E_i$ share the same lepton number as does the SM charged 
lepton, here represented as an electron.

We begin by first considering the $Q_{em}=-1$ states which form a $4\times 4$ mass matrix, ${\cal E}^cM_{-1}{\cal E}$, where ${\cal E}=(e,E_1,E_2,E_3)^T$ and 
\begin{equation}
\sqrt 2 M_{-1}=~\begin{pmatrix} y_2v_2 & 0 & B &C \\ D & y_3u_2 &y_2v_1 & y_2v_4 \\ 0 & y_2v_2& y_4 u_4 &-y_4u_3\\F & -y_3u_1 &y_2v_5 & y_2v_3\\ \end{pmatrix}\,.
\end{equation} 
Similarly, for the $Q_{em}=0$ fermions, the various mass terms now lead to the $5\times 5$ {\it Dirac} mass matrix, ${\cal N}^cM_0^D{\cal N}$, where the neutral 
fermion basis is chosen to be as ${\cal N}={\nu,S_1,N_1,N_2,S_2}$ and 
\begin{equation}
\sqrt 2 M_0^D=~\begin{pmatrix} y_2v_1 & y_2v_5& 0 & -B &0 \\ y_2v_4 & y_2v_3 &0 & -C & 0 \\ -D &-F & -y_3u_2& y_2 v_2 &y_3u_1\\0 & 0 & y_2v_1 &-y_4u_4 & y_2v_5\\0 & 0 &y_2v_4 & y_4u_3 &y_2v_3\\ \end{pmatrix}\,.
\end{equation} 
In both of these matrices, the shared entries represented by the upper case Roman letters $B,C,D,F$ are somewhat special in that they arise from the rather small, $Q_D\neq 0$ vevs, 
$x_i\lsim 1$ GeV, and play an essential role in the phenomenology of the new states appearing in this setup (\ie, particle decays) but can be essentially neglected when discussing the (Dirac) masses 
themselves to a very good approximation. Note that the lepton number violating vevs, $x_{2,4}$, do {\it not} appear here. In particular, we have that
\begin{equation}
B=-y_4x_6,~~~C=y_4x_5,~~~D=-y_3x_3,~~~F=y_3x_1\,.
\end{equation} 
We will return to the {\it Majorana} mass terms that are generated by the remaining $Q_D\neq 0$, $\Delta L=2$ vevs, $x_{2,4}$, in the next Section below.

There are many observations that one can make about the overall structure of the matrices $M_{-1}$ and $M_0^D$ with respect to their apparent sub-components that result from 
both the locations of null entries as well as the hierarchy of the various vevs which make their appearances within them. In the case of the $Q_{em}=-1$ fermions, we first observe that, 
unsurprisingly, in the absence of the small $Q_D\neq 0$ vevs, $x_i$, the electron is `isolated' and does not mix with the remaining VL $E_i$ states since, unlike SM fields, they all have 
$Q_D=-1$. However, once these small vevs are turned on, $e-E_i$ mixing of a generally chiral nature is induced allowing for the 2-body decays $E_i\to e+$DP. As has frequently been 
discussed and is by now well-known
\cite{Rizzo:2018vlb,Rueter:2019wdf,Rueter:2020qhf,Wojcik:2020wgm,Rizzo:2021lob,Rizzo:2022qan,Wojcik:2022rtk,Rizzo:2022jti,Rizzo:2022lpm,Wojcik:2022woa,Rizzo:2023qbj,Wojcik:2023ggt}, 
due to the enhanced coupling of the DP's longitudinal component (or the $O(1)$ Yukawa coupling of the equivalent Goldstone boson\cite{GBET}) arising from the large ratio of the mass of the 
PM lepton to that of the DP, this is the dominant PM decay mode over essentially all of the entire model 
parameter space and provides the production/decay signature for these particles at colliders\cite{Rizzo:2022qan}, \ie, an oppositely charged, same-flavor charged lepton pair 
plus MET, assuming the DP decays to DM or is sufficiently long-lived.  In any case, naively, we would roughly expect the size of this SM-PM mixing to be on the order of the ratio of the 
relevant vevs/mass scales, \ie, $\sim 1$ GeV/$\sim10$ TeV $\sim 10^{-4} \sim \epsilon$, or so as has been noted in previous works, which is essentially just the inverse of the large mass 
ratio enhancement discussed above.
At the opposite end of the mass spectrum, we see that when we turn off the $\sim 100$ GeV, $SU(2)_L$ violating vevs, $v_{1,2,4}$ as well as $u_1$, one finds that $E_1$ no longer mixes 
with $E_{2,3}$ while both the left- and right-handed components of these later two states can still mix significantly depending upon the relative sizes of the remaining large vevs. The masses 
and mixings among these three states will depend upon the values and mass ordering among the set of large vevs which is clearly correlated with the pattern of gauge symmetry 
breaking within this setup but in any case the mixing of $E_1$ with $E_{2,3}$ is expected to be of order $\sim 100$ GeV$/\sim10$ TeV $\sim 10^{-2}$. However, the corresponding mixing between 
the remaining states $E_{2,3}$ is likely to be $\sim O(1)$ for both helicities unless special hierarchies amongst the various vevs are preferred by the gauge symmetry breaking patterns to be 
discussed below.

In the case of the Dirac masses for the neutral fermions, we find a somewhat similar pattern outside of some essential differences due to, amongst other things, there now being 5 fields that mix
instead of 4 and that the SM $\nu$ as well as $S_1$ fields both have $Q_D=0$ here. We also observe that the $SU(2)_{L,R}$ partners $(N_1,E_1)$ and $(N_2,E_2)$ are (separately) 
degenerate in the absence of mixing as might be expected. We similarly note that $S_2,E_3$, which are both $SU(2)_{L,R}$ singlets, are also found to be degenerate in this same limit. 
In addition, as we will discuss in the next Section, Majorana mass terms are likely also present amongst some of these neutral fields, inducing further mixings beyond those discussed below, 
which arise solely from the lepton-number violating subset of the $Q_D$-violating vevs, $x_{2,4}$. Here we observe that the most essential 
difference, as far as the Dirac mass terms are concerned, with respect to the $Q_{em}=-1$ case, is that there exists an additional 
fermion, $S_1$, which, together with the SM $\nu$, is also a $Q_D=0$ fermion. In the limit that the $Q_D\neq 0$ vevs can be ignored, this $5\times 5$ mass matrix breaks down into a 
$2\times 2$ block, $M_{\nu,S_1}$, for the $Q_D=0$ fields, $\nu,~S_1$, as well as a $3\times 3$ block for the $Q_D=-1$ fields $N_{1,2},S_2$; recall that $N_{1,2}$ share isodoublets with 
the previously discussed electrically charged PM fields $E_{1,2}$. 

Our first observation is, that apart from an overall Yukawa coupling, the submatrix $M_{\nu,S_1}$ is identical to $M_{-1/3}$ 
and so can be diagonalized by the same rotations as described above. This mixing allows for the $2_L2_R$ isosinglet to have suppressed decays through both SM as well as 
LRM-like interactions. The $3\times 3$ submatrix is, essentially, apart from some signs, observed to be the transpose/hermitian conjugate of 
the $3\times 3$ submatrix appearing in $M_{-1}$ above and, like there, if the $SU(2)_L$-violating vevs are turned off, one sees that the corresponding field $N_1$ no longer mixes with 
the remaining states $N_2,S_2$. This implies that, similar to the case of $E_1$ above, the mixing between $N_1$ and both $N_2,S_2$ will be of order $\sim 100$ GeV$/\sim10$ TeV 
$\sim 10^{-2}$. We again note that due to the $SU(2)_{L,R}$ symmetries, the pairs of states $N_1,~E_1$ and $N_2,~E_2$ are degenerate prior to both their mixings with the other states as well as 
loop radiative corrections; this is also observed to be true for the states $S_2, ~E_3$ due to the extended $SU(3)_{L,R,'}$ symmetries. This implies that that 2-body decays, such as,\eg, 
$E_1\to N_1W_{SM}$, even if kinematically allowed, will more than likely be highly suppressed by the very small phase space availability; even less likely are any on-shell, 2-body decays 
via $W_{R}^\pm$ since these new gauge states are substantially more massive. The parallel with the $Q_{em}=-1$ states is further strengthened by the presence of the $Q_D\neq 0$ vevs that allow 
for both $N_{1,2}$ to directly mix in a chiral manner with $\nu, S_1$ while $S_2$ mixes only indirectly with them and must first mix with $N_{1,2}$ beforehand. As in the $Q_{em}=-1$ case, 
we might expect this $Q_D$-violation induced mixing to be $\sim 1$ GeV/$\sim10$ TeV $\sim 10^{-4} \sim \epsilon$, allowing for the presence of the dominant decay modes 
$(N_{1,2},S_2) \to (\nu,S_1) +$DP, with the DP in most cases will likely appear simply as MET.
As will be seen below, $S_1$ can be produced through the production and decay of $U_L$, the mixing of $U_L$ with the SM $W$, the mixing of $S_1$ 
with the SM $\nu$ or some combination of these mechanisms

\section{Neutral Fermion Majorana Masses}

As noted in the previous Section, in the neutral fermion sector, all of the $Q_D=0$ vevs produce only Dirac mass terms but two of the $Q_D$-violating vevs, $x_{2,4}$, which correspond 
to scalar fields of opposite $Q_D$ charge from those sourcing the other $x_i$, can generate $\Delta L=2$ Majorana mass terms. Since $H_{2,3}$ are the scalar analogs of $l,l^c$ in that they 
transform under $3_c3_L3'3_R$ in exactly the same manner, we observe that the `centeral' neutral member of these multiplets are distinctive in this manner as can be seen from 
Table~\ref{vevtab}. However, these vevs only result in the following rather simple pair of Majorana mass terms linking 4 of the 5 neutral fermion fields which can then be used to define the 
corresponding left- and right-handed Majorana mass matrices, $M_{L,R}$: 
\begin{equation}
y_3\Big(\nu S_2-N_1S_1 \Big)\frac{x_2}{\sqrt 2} +y_4\Big(\nu^cS_2^c-N_1^cS_1^c\Big)\frac{x_4}{\sqrt 2} \to ~ M_L+M_R\,,
\end{equation} 
which are in some sense, almost conjugates of each other; note that $N_2$ is not involved in this interaction. These matrices, when combined with $M_0^D$, will then form a more general 
$10\times 10$ Majorana mass matrix. With the perhaps naive expectations that $x_2 \sim x_4$ and $y_3 \sim y_4$ 
we might expect then that $M_L \simeq M_R$. Needless to say, in comparison to most of the entries in the neutral Dirac mass matrix given above, these Majorana mass matrices are quite 
sparse, having non-zero elements that are all quite small, $\lsim 1$ GeV, as the relevant vevs are both $U(1)_D$-violating as well as being lepton-number violating. As a result they will have little 
influence on the heavy neutral lepton states (except for the possibility of making them pseudo-Dirac with extremely small fractional mass splittings in some cases \cite{quasi} as we encountered in our 
previous study based on the Pati-Salam setup\cite{Rizzo:2023qbj}) and they are not capable of 
explaining the observed light neutrino masses without further extensions of this setup as has been noted elsewhere in the Quartification literature\cite{quart}.

\section{Non-hermitian Gauge Boson Masses and Mixings}

Apart from QCD, the gauge boson sector of the $3_c3_L3'3_R$ Quartification setup is rather complex with a total of 24 gauge fields of which 6 are hermitian while the remainder form 9 
pairs of complex, \ie, non-hermitian fields including the SM $W_L^\pm$ as well as the corresponding $W_R^\pm$, familiar from the LRM. It should be noted that not all of these non-hermitian fields 
carry electric charge, \ie, have $Q_{em}=\pm 1$. In this section, the masses of these non-hermitian gauge bosons (NHGB) will be discussed and, as we'll see, an additional Higgs scalar, 
$\Omega$, beyond the 3 discussed above, the $H_i$, that are responsible for fermion mass generation, will need to be introduced to satisfy our model building assumptions. The 
corresponding discussion for the 6 hermitian fields will be the subject of the next Section. Except for rare circumstances and in the absence of any large mixing effects, the couplings of the 
NHGB to the various fermions will generally be chiral, a notable exception being, \eg, that of the SM[LRM] $W_L[W_R]$ to the $(N_1,E_1)^T[(N_2,E_2)^T]$ doublet which is vector-like.

Before beginning, the first observation to make is that, apart from $W_L^\pm$, all of these 9 NHGB need to be heavy, likely in excess of several TeV, to avoid the many LHC 
searches\cite{wpsearch,ATLAS:2022jsi,CMS:2023ooo} and so must have their masses generated by combinations of the vevs $v_{3,5}, u_{2-4}$ above as well as possibly by others of a 
similar magnitude. A second observation is that the 3 NHGB arising from 
$SU(3)'$ all carry different values $Q_D \neq 0$ (as well as different values of $Q_{em}$) and so, in the limit that $U(1)_D$ remains unbroken, they will not mix with the remaining 6 fields or 
even with each other.  A third observation is, given the definition of the electric charge 
in terms of the gauge group generators above, that the 3 pairs of NHGB associated with each of the $SU(3)_a$ gauge will consist of 2 pairs of fields with $Q_{em}=\pm 1$, which we will 
generically refer to as $W_a^\pm,U_a^\pm$, while the remaining pair of NHGB is electrically neutral, which we will refer to as $V_a^{0(\dagger)}$. These two observations, when combined, 
tell us that the full $9\times 9$ mass matrix for the NHGB in the limit of $U(1)_D$ conservation will consist of 3 blocks: a diagonal $3\times 3$ block for the 3, $Q_D \neq 0$, $SU(3)'$ fields, a 
$2\times 2$ block for the $Q_D=0$, neutral fields $V_{L,R}$, and the remaining $4\times 4$ block for the electrically charged, $Q_D=0$ fields $W_{L,R},U_{L,R}$, respectively. Lastly, it 
is useful to be reminded about what the roles are for the three classes of gauge bosons, $W_a,U_a$ and $V_a$, in the sense of which fermion fields, that here lie in $\bf 3$'s or 
$\bf{\bar 3}$'s of $SU(3)_a$, are connected to each other by them. If we label the three triplet fermions a $(f_1,f_2,f_3)^T$, then $W$ connects $f_1-f_2$ while $V$ connects $f_1-f_3$ 
and $U$ connects $f_2-f_3$. Thus, \eg, $W_L$ connects $(\nu,e)$ and $(d,u)$ in the familar way, while $V_L$ connects $(\nu,S_1)$ and $(d,h)$ and $U_L$ connects $(e,S_1)$ and $(u,h)$, 
respectively. In practice, amongst other things, this implies that $W_a$ and $U_a$ will have opposite values of $Q_{em}$ so that, \eg, $W_L$ mixes with $U_L^\dagger$ and 
not with $U_L$.  Finally, with this notation, one finds that  $Q_D(W',V',U')=(1,-1,-2)$ and, given that $Q_{em}(V')=0$, we see that this NHGB plays essentially the same role (in the leptonic 
sector) as did the $W_I$ gauge field in the $G_{Dark}=SU(2)_I\times U(1)_{Y_I}$ model encountered in several of our earlier works, while the $W'$ acts in a similar manner being its electrically 
 charged partner. The gauge field $U'$, on the other hand, in the absence of any mixing, will only connect pairs of states with non-zero values of $Q_D$ and so will not easily be produced 
 or probed in a simple fashion, especially at a hadron collider while future searches in conventional $W'$ channels at FCC-hh may reach as high as $\sim 40$ TeV \cite{wpsearch,ATLAS:2022jsi,Rizzo:2014xma,Nemevsek:2023hwx}.  
 
 It is interesting to note that since the $W'$ and $U'$ gauge bosons carry no-zero values of both $Q_D$ and $Q_{em}$, they 
 too can act as PM fields, something we had not previously encountered but is an obvious result of the embedding of $U(1)_D$ within $SU(3)'$ and that fact that $Q_{em}$ also depends upon 
 the diagonal $SU(3)'$ generators.
 Another interesting feature of these NHGB states is also to be observed: while $W_{L,R},U_{L,R}$ and $V_{L,R}$ will clearly carry zero lepton number, $L=0$, this is {\it not} generally true for 
 the $SU(3)'$ NHGBs. In fact, one finds that $W',U'$ carry $|L|=2$ while $V'$ remains an $L=0$ state. We will return to this issue below when we consider the mixing induced by the small 
 $x_{2,4} \neq 0$ vevs which we have seen above generate $\Delta L=2$ Majorana mass terms among the new neutral fermions..

With this preparation we will now consider the $4\times 4$ ($M^2_{44}$), $2\times 2$ ($M^2_{22}$) and $3\times 3$ ($M^2_{33}$) blocks of the full $9\times 9$ NHGB mass-squared matrix 
that are obtained in the approximate limit that we can ignore the $\lsim 1 $ GeV, $Q_D \neq 0$ vevs; we will then return and discuss the perturbing effects that these additional vevs will have. As a 
first step, we will assume that the only Higgs vevs that are relevant are those that are employed above to generate the various fermion masses and whose properties are already given 
in Table~\ref{vevtab}. In the $(W_L,U_L^\dagger,W_R,U_R^\dagger)${\footnote {Recall that, \eg, $W_L$ and $U_L^\dagger$ carry the same electric charge.}} basis we find that 
\begin{equation}
 M^2_{44}=\frac{1}{4}~\begin{pmatrix} g_L^2(v_1^2+v_2^2+v_4^2+u_1^2) &g_L^2(v_1v_5+v_3v_4+u_1u_2) & 2g_Lg_Rv_1v_2 & 2g_Lg_Rv_2v_4\\
g_L^2(v_1v_5+v_3v_4+u_1u_2) & g_L^2(v_3^2+v_5^2+u_2^2) & 2 g_Lg_R v_2v_5 & 2g_Lg_Rv_2v_3\\
2g_Lg_Rv_1v_2 &2 g_Lg_R v_2v_5 &g_R^2(v_1^2+v_2^2+v_5^2) & g_R^2(v_1v_4+v_3v_5+u_3u_4)\\
2g_Lg_Rv_2v_4 &2g_Lg_Rv_2v_3 &g_R^2(v_1v_4+v_3v_5+u_3u_4) & g_R^2(v_3^2+v_4^2+u_4^2)\\ \end{pmatrix}\,,
\end{equation} 
whereas, in the $(V_L,V_R)$ basis we obtain 
\begin{equation}
M^2_{22}=\frac{1}{4}~\begin{pmatrix} g_L^2X_L & 2g_Lg_R(v_1v_3+v_4v_5)\\ 2g_Lg_R(v_1v_3+v_4v_5) &g_R^2X_R\\ \end{pmatrix}\,,
\end{equation} 
where $X_L=v_1^2+v_3^2+v_4^2+v_5^2+u_1^2+u_2^2$ and $X_R=v_1^2+v_3^2+v_4^2+v_5^2+u_3^2+u_4^2$. Note that $V_L-U_L$ have their masses split by electroweak scale vevs 
only as they form an $SU(2)_L$ doublet. Finally, in the $(W',U'^\dagger,V')$ basis we find 
\begin{equation}
M^2_{33}=\frac{g'^2}{4} X'~\begin{pmatrix} 1 & 0 & 0\\ 0 & 0 & 0\\ 0 & 0 & 1\\ \end{pmatrix}\,,
\end{equation} 
where $X'=u_1^2+u_2^2+u_3^2+u_4^2$. 

There are many things to observe about these results, beginning with the most obvious ones arising from $M^2_{33}$: $U'$ is massless and 
$W',V'$ are degenerate. The masslessness of $U'$, alluded to above, is the result of the requirement of preserving $U(1)_{em}$ as well as $U(1)_D$ down to low mass scales so that the 
Higgs in the (anti-)triplet representations only break $SU(3)'$ down to $SU(2)'$. Clearly $U'$ must be a rather massive field with $M_{U'} \gsim $ a few TeV or so and thus we must employ 
an additional Higgs scalar with vevs beyond those found in the $H_i$ which transforms differently under $SU(3)'$. The simplest possibility is is to employ an $SU(3)'$ octet, \ie, $\Omega$, that is an  
$(1,1,8,1)$ representation under $3_c3_L3'3_R$, which will not influence the fermion masses discussed above, will not contribute to the masses of the Hermitian gauge bosons that we 
will discuss in the next Section, nor to the masses of the other NHGB which are all $SU(3)'$ singlets. Note that this additional scalar representation represents an explicit breaking of the 
apparent symmetry $Z_4$ among the gauge fields and fermions dictated by anomaly freedom and the various $[SU(3)]^4$ group factors but, as we've noted above, we've {\it not} assumed 
that such a symmetry exists as part of our present discussion. We will return to this issue later below. As is well-known, the vevs of an $SU(3)$ octet acting on (anti-)triplet representations 
can be expressed as 
\begin{equation}
<\Omega>~ \sim ~\begin{pmatrix} w_1 & 0 & 0\\ 0 & w_2 & 0\\ 0 & 0 & w_3\\ \end{pmatrix}\,,
\end{equation} 
subject to the constraint that the sum of these vevs satisfies $\sum_i w_i=0$ due to the tracelessness of the octet. This results in a shift, $\Delta M^2_{33}$, in $M^2_{33}$ that is given by 
\begin{equation}
\Delta M^2_{33}=\frac{g'^2}{8}~\begin{pmatrix} (w_2-w_1)^2 & 0 & 0\\ 0 & (w_3-w_2)^2 & 0\\ 0 & 0 & (w_3-w_1)^2\\ \end{pmatrix}\,,
\end{equation} 
thus resolving our problem with the $U'$ mass (under the assumption that the vevs $w_i \gsim$ a few TeV or so and that $w_3\neq w_2$) and also simultaneously removes the $W'-V'$ 
mass degeneracy, not that this was a problem in any way. From this result it is easy to imagine that the $U'$ may be the lightest of the three $SU(3)'$ NHGB.

Turning now to $M^2_{22}$, we see both diagonal entries involve the squares of large vevs whereas the off-diagonal terms also involve the weak-scale vevs, $v_{1,4}$, so that this 
matrix is diagonalized via the small angle 
\begin{equation}
\tan 2\phi_{22} =\frac{2\kappa(v_1v_3+v_4v_5)}{X_L-\kappa^2X_R}\,
\end{equation} 
where $\kappa=g_R/g_L\simeq 1$, and which we might thus expect to be of order $\sim 10^{-2}$ or so. Again, with $\kappa \simeq 1$, the masses resulting for both of the eigenstates, $V_{1,2}$, 
will clearly lie in the range of at least several TeV.

Lastly, $M^2_{44}$ is seen to be the most complex of these 3 sets of mass-squared blocks but one immediate observation is that the limit that the electroweak scale vevs are neglected, not only 
is SM $W_L$ massless but its mixings with the remaining 3 states, who dominantly obtain their masses from the larger vevs, will all vanish. Looking a bit closer we see that whereas we 
might expect that the mixing of the $W_L$ with $W_R,U_R^\dagger$ to be of order $ \sim (100~ {\rm GeV}/10 ~{\rm TeV})^2 \sim 10^{-4}$, its mixing with $U_L^\dagger$ is potentially much 
larger, of order $ \sim 100~ {\rm GeV}/10 ~{\rm TeV}  \sim 10^{-2}$, which {\it could} lead to significant phenomenological consequences as it, \eg, `depletes' a small amount of the $\bar u d$-type 
coupling into a $\bar u h$-type coupling. However, since an {\it identical} effect occurs in the leptonic sector (since gauge boson mixing has nothing to do with the fermion sector at tree level) 
in that the $\bar \nu e$-type coupling is also diverted into an $\bar S_1 e$-type coupling, the effect of this tree-level $W_L-U_L^\dagger$ mixing may be more subtle than it first 
appears, requiring a consistent treatment since it is a `universal' effect experienced by both the quark and lepton sectors. In the other parts of this matrix, we see the $W_R-U_R^\dagger$ mixing 
can be $O(1)$ since it involves only the large vevs ({\it assuming} that there is also no significant hierarchy amongst these large vevs which may or may not be the case depending upon how 
the guage symmetries are broken) while the mixing of these two fields with $U_L^\dagger$ is also seen to naively be relatively small, of $O(10^{-2})$. In a manner similar to $W_L-W_R$ mixing, 
we see that $U_L-U_R$  mixing may also roughly lie at the level of $10^{-2}$.

Since $W_R-U_R^\dagger$ mixing is allowed to be significant $\sim O(1)$, they will likely both couple to the $\bar ud+{\rm h.c}$ initial state though the lighter of the two mass eigenstates, $\tilde W$, 
would have the largest production cross section but would now have two distinct leptonic decay modes. Since SM neutrinos are likely Dirac in the present setup, $\tilde W$ will decay into both 
the $e\nu \to e+$MET as well as the $eS_1^c\to eejj$ final states with somewhat comparable branching fractions, allowing for phase space. This implies that a signal may be observable in more 
than one search mode\cite{wpsearch,ATLAS:2022jsi}.

It should be noted that whereas $W_L-U_L^\dagger$ mixing will result in a small downward shift in the SM $W$ mass, the SM $Z$ mass, as we will see below, is also pushed downward due to 
a similar mixing. However, we note that the combination of these two effects, at tree level, will most likely not provide any explanation for the apparent upward shift in the $W$ mass, relative to 
that of the $Z$,  as measured by CDFII\cite{CDF:2022hxs}.

Overall, we see that apart from the (approximate) SM $W_L$ mass eigenstate, all of the NHGB masses will lie at the (at least) several TeV scale but their specific mass spectrum being quite 
sensitive to the various potentially large mixings among these states as well as the specific ordering of the values of the multiple high-scale vevs $v_{3,5},u_{2-4}$ as well as $w_{1-3}$. 

Finally, we now discuss how this picture of the NHGB masses and mixings is altered when we turn on the small $Q_D$-violating vevs, $x_i$, appearing in the Higgs fields $H_{2,3}$ above. 
Clearly, some previously unmixed states with the same value of $Q_{em}$ can now mix but the most important effect is most obviously the specific induced mixing between states of 
{\it different} $Q_D$ that were forbidden to mix previously. Mixings can now also occur between states with differing lepton number as the $x_{2,4}$ violate this quantity as was noted earlier. 
Specifically, this also means that, \eg, there is now a small induced mixing of $V'$ {\it and} $V'^\dagger$ with both $V_{L,R}$ and $V_{L,R}^\dagger$ (all of which mixings are lepton 
number conserving).  There are also correspondingly small mixings of the $W'$ with the fields $W_{L,R}$, \etc, with specific sets 
of $x_i$ participating, and, furthermore, there is now an induced mixing between the two $SU(3)'$ fields $W'-U'^\dagger$ at second order, $\sim x_ix_j$.  We note that there is 
{\it no} mixing between the $U'$ and either $W_{L,R}$ or $U_{L,R}$ to leading (\ie, linear) order in $x_i$ as all of the vevs are $|\Delta Q_D|=1$ and these states differ by 2 units of $Q_D$ as 
well as by lepton number. However, still at leading order in the $x_i$, there is such mixing between the $U_{L,R}$ and the $W'$.  One might expect that, 
roughly speaking, the size of these typical mixings to be or order, \eg, $\phi_{W_{L,R}W'}\sim x_{2,4}/(v_{3,5},u_{2-4}) \sim 10^{-(4-5)}$ or so. Phenomenologically, while these tiny mixings 
have very little effect on any of the NHGB masses, they will now allow (via the $x_{2,4}$ $\Delta L=2$ vevs) for new decay paths, \eg, $W'\to W_{L,R}+D$ assuming that the $W'$ is more massive 
than the $W_R$.  As in the well known from the example of PM decay induced by a similar type of mixing, the 
longitudinal coupling of the DP, \ie, $D$, (and also that of the SM/LRM $W_{L,R}$) is enhanced by a double ratio of large masses, $M^2_{W'}/(M_{W_{L,R}}M_{\rm D} )\sim 10^4$ 
or so, which is likely sufficient to overcome the correspondingly expected tiny mixing angle suppression so that this may become one of (if not solely) the dominant $W'$ decay mode(s); in 
this example, the corresponding (single) production signature at a collider would then be just $W+$MET for which, \eg, LHC searches exist\cite{ATLAS:2018nda} assuming these $W'$ states 
are kinematically accessible, which is one of the issues that we will return to below. Explicitly, to leading order in these large mass ratios, we obtain, \eg, the following expression this partial width:
\begin{equation}
\Gamma(W'\to W_L+{\rm D})\simeq \frac{\alpha_D M_{W'}}{48}~\Big[\phi_{W_LW'}\frac{M^2_{W'}}{M_{W_L}M_{\rm D}}\Big]^2\,
\end{equation} 
where $\alpha_D=g_D^2/4\pi$ with $g_D$ being the DP's $U(1)_D$ gauge coupling, $\sim g'$ as we will find in the next Section below, and where we expect the expression in the square 
bracket to be $O(1)$.  

While some of the NHGB may be easily produced a hadron collider, this will not be true for all of them as a subset will only interact with $Q_D\neq 0$ sector fields and not directly with those of 
the SM except via mixing which leads us directly to the some of the discussions in later Sections. However, the same mixing that allows for the $W'\to W_L+D$ decay process can also be 
used to singly produce a $W'$ together with a DP at a hadron collider via, \eg, an off-shell transverse $W_L$ exchange in the $s$-channel. In such a case, one would find that the cross section 
scales as the square of $\phi_{W_LW'}\frac{M_{W'}}{M_{\rm D}}$, which is seen to be a factor of $\frac{M_{W_L}}{M_{W'}} \sim 10^{-2}$ smaller in amplitude than that appearing in the square bracket 
above and which was presumed to be $O(1)$. This would then imply that the cross section for this $W'$ production process is relatively suppressed by a factor of roughly $\sim 10^{-4}$ and 
so would be `difficult' to observe at best unless the $W'$ were not to bee too massive.

\section{Hermitian Gauge Boson Masses and Mixings}

In some ways, the mass-squared matrix for the six Hermitian gauge bosons (HGB), \ie, $W_{3a},W_{8a},  (a=L,R,')$, is more complex than that encountered above in the NHGB case although 
we have several guideposts thanks to the expected hierarchy of symmetry breaking scales. For example, in the limit that we can ignore the $\sim 1$ GeV, $Q_D$-violating vevs, this 
mass-squared matrix will have  two zero eigenvalues corresponding to both the DP as well as the usual SM photon and, furthermore, in the limit of vanishing $SU(2)_L$-violating vevs, the SM 
$Z$ must also be massless. Finding a convenient and useful basis for this matrix is, however, somewhat non-trivial and clearly the above set of HGB,  $W_{3a},W_{8a}$, is itself not a very 
useful choice in this regard. Given the various Higgs multiplets introduced above, we have sufficient freedom to generate all of the HGB masses as we will see below. 

One obvious observation is that, due to the definition of $Q_D$ and the fact that $U(1)_D$ remains unbroken down to low mass scales, 
we can decompose the $SU(3)'$ terms corresponding to the diagonal generators appearing in that part of the 
covariant derivative (dropping Lorentz indices) as the two orthogonal combinations{\footnote {This normalization has been chosen so that, as usual, $Tr [G_i]^2=1/2$, where the $G_i$ 
are the combinations of the $SU(3)'$ generators appearing in the brackets, obtained for either the fundamental ${\bf 3}$ or anti-fundamental $\bf{\bar 3}$ representation.}}
\begin{equation}
g'(T_3'W_3'+T_8'W_8') =g'~\frac{\sqrt 3}{2} ~[T_3'+T_8'/\sqrt 3]~W_+'+\frac{g'}{2}~[T_3'-\sqrt 3 T_8']~\tilde D\,,
\end{equation} 
with the definitions
\begin{equation}
W_+'=\frac{\sqrt 3}{2} ~[W_3'+W_8'/\sqrt 3], ~~~~\tilde D=\frac{1}{2}~[W_3'-\sqrt 3 W_8']\,.
\end{equation} 
Here we recognize the expression in the square bracket associated with gauge field $\tilde D$ as just equal to $Q_D$ (up to a possible sign) such that {\it if}  
$\tilde D$ were to be identified with the DP, $D$, with this chosen normalization the usual $U(1)_D$ 
gauge coupling must then be just $g_D=g'/2=g's_I$. In the limit where we can neglect the contributions arising from the $Q_D \neq 0$ vevs, $x_i$, as a good approximation, we can simply omit 
the appearance of this DP term in the covariant derivative when constructing the HGB mass-squared matrix which, effectively,  now becomes only $5\times 5$ since $\tilde D$ neither has a mass 
or any off-diagonal mixing terms with the other HGB states. When combined with the  
quantum numbers of the various Higgs fields in Table~\ref{vevtab}, this result motivates us to consider the following useful (but not always physically intuitive) basis for these 
neutral gauge fields: 
\begin{equation}
A_a = g_a (W_{3a}+W_{8a}/\sqrt {3}),~~~~ B_a= g_a \frac{2}{\sqrt 3}W_{8a}\,,
\end{equation} 
with, as usual, $a=L,R,'$. It is important to note that the gauge couplings have been absorbed into these definitions; note also that $A' \sim W_+'$ as defined in the 
equation above, apart from a normalization factor. Interestingly, in this basis, instead of the $SU(3)'$ part of the covariant derivative decomposition above, we find that, \eg, 
\begin{equation}
g'(T_3'W_3'+T_8'W_8') =T_3' ~A'-\frac{1}{2}~[T_3'-\sqrt 3 T_8']~B'\,,
\end{equation} 
and similarly for $a=L,R$. Note that, due to the possible $Q_D$ sign assignment ambiguity, apart from an absorbed gauge coupling, here $B'$ might {\it alternatively} be identified as the 
DP, $D$, again with $g_D=g'/2$ as before{\footnote {One could instead just redefine $D=-B'$ and ignore this sign issue altogether.}} More generally, we can define the $B_a$ fields as 
coupling to the generalized charges $Q_a$ with $Q'=-Q_D$. In this basis, neglecting the $Q_D\neq 0$ vevs, $x_i$, for the moment as we did above, the $B'$ field decouples and remains 
massless and so the general $6\times 6$ HGB mass squared matrix again reduces to one which is $5\times 5$ by the earlier argument.

In this $A_L,A_R,B_L,B_R,A'$ basis, we can write this truncated (the $B'$ or DP now being omitted) $5\times 5$ HGB mass squared matrix as 
\begin{equation}
 M^2_{HGB-5}=\frac{1}{4}~\begin{pmatrix} v_1^2+v_2^2+v_4^2+u_1^2 & -v_1^2-v_2^2 & -v_2^2 & v_2^2+v_4^2 & -u_1^2\\
 -v_1^2-v_2^2 & v_1^2+v_2^2+v_5^2+u_3^2 & v_2^2+v_5^2 & -v_2^2 & -u_3^2\\
 -v_2^2 & v_2^2+v_5^2 & v_2^2+v_3^2+v_5^2+u_2^2 & -v_2^2-v_3^2 & u_2^2 \\
 v_2^2+v_4^2 & -v_2^2 & -v_2^2-v_3^2 & v_2^2+v_3^2+v_4^2+u_4^2 & u_4^2 \\
  -u_1^2 & -u_3^2 & u_2^2 & u_4^2 & U\\
\end{pmatrix}\,,
\end{equation} 
where we have defined the quantity $U=\sum_i u_i^2$. It is important to remember when viewing this matrix that the gauge couplings do not appear as they have been absorbed into the 
definitions of the set of the gauge fields, $A_a,B_a$.  Here we observe that the first row and column consists solely of combinations of the $SU(2)_L$-breaking vevs so that, in the limit  
where the squares of the electroweak-scale breaking can be neglected and only the mult-TeV scale vevs are relevant, the $A_L$ gauge boson will decouple as `massless'. In such a case, 
this matrix can 
effectively be further truncated to the much simpler, lower right-hand, $4\times 4$ block with the vevs $v_{1,2,4},u_1$ all set to zero in this approximation. Explicitly, this matrix 
is found to be given as 
\begin{equation}
 M^2_{HGB-4}\simeq \frac{1}{4}~\begin{pmatrix} 
 v_5^2+u_3^2 & v_5^2 & 0 & -u_3^2\\
 v_5^2 &v_3^2+v_5^2+u_2^2 & -v_3^2 & u_2^2 \\
 0 & -v_3^2 & v_3^2+u_4^2 & u_4^2 \\
 -u_3^2 & u_2^2 & u_4^2 & u_2^2+u_3^2+u_4^2\\
\end{pmatrix}\,.
\end{equation} 
In this same limit, we should expect this truncated $4\times 4$ mass-squared matrix to have one null eigenvalue whose associated field, which well call ${\cal H}$, which together with $A_L$, will 
form the familiar $W_3,B_Y$ ones of the SM.  In fact, unsurprisingly, it is easily seen that in this truncated basis (and here taking $g_L=g_R=g'=g$ as is frequently done in the Quartification 
literature\cite{quart} for clarity of presentation except where noted), that ${\cal H}\sim (1,-1,-1,1)^T$, \ie, the combination of fields 
$\sim (A_R-B_L-B_R+A')$,  forms the null eigenvector corresponding to a massless HGB state{\footnote {Importantly, we recall that each of the remaining three mass 
eigenstates/eigenvectors must be orthogonal to this state as well as to each other.}}.  Interestingly, if we return to the $5\times 5$ matrix above, also including the additional electroweak scale 
vevs, then in this more general case we would, perhaps unsurprisingly, find that the corresponding null eigenvector is instead $\sim (1,1,-1,-1,1)^T$, on it's way to being identified with 
the photon given the definition of $Q_{em}$.  Returning now to the $4\times 4$ case at hand, we note that, since all these vevs are large, we might expect that these 4 states, $A_R,B_{L,R}$ 
and $A'$, will in general mix together rather strongly unless the vevs have some sort of associated hierarchy about which we have no {\it a priori} knowledge; this implies that general 
expression for all the mass eigenvalues and eigenstates in terms of these vevs would be hardly enlightening.  

One possibility is to make a choice of basis based on familiar dynamics 
which suggests that an examination of these HGB states in a phenomenological basis.  Since, from the large vev perspective, three of these HGB are clearly either massless or close to 
being massless eigenstates to a fairly good approximation, one might consider employing, \eg, $\gamma, D, Z_{SM}, Z_R, \tilde B, A'$, where $Z_R$ is the familiar and 
well-studied $Z'$ of the LRM and $\tilde B \sim B_L-B_R \sim g_LW_{8L}-g_RW_{8R}$. On the other hand, it is may be more useful to explore the nature of this matrix under variously 
motivated assumptions about the vevs themselves, \eg, one might expect that the product $3_L3_R$ breaks at a very high scale to $2_L2_R1_L1_R \to 2_L2_R1_{L+R}$, similar in nature 
to the LRM, although a piece of both $A',B'$ will also be necessary to form the familiar $1_{B-L}$ gauge group factor as can be seen from the definition of $Q_{em}$ in terms of the set of 
diagonal generators. Note that $\tilde B$ is just the orthogonal combination to that which appears in the gauge field for $U(1)_{B-L}$, and will 
be discussed in more detail below. We recall that in the well known Trinification model\cite{trinif}  limit this field is seen to be just the result of the reorganization and/or diagonal breaking of the product 
$1_L1_R \to 1_{B-L}1_{\tilde B}$. 

The field $\tilde B$, in fact, provides an excellent (at first toy) case in point for both studying the influence of vev hierarchies as well as a 
phenomenological motivation for choosing a particular HGB mass eigenstate basis as we see that the large vev, $v_3$, preserves, \eg, in the generalized charge notation above, the combination 
$Q_L/2+Q_R/2$ but breaks the difference $Q_L/2-Q_R/2$. To clarify for demonstration purposes, consider for simplicity rewriting only the relevant pieces in the covariant derivative and neglecting 
the contributions from the other gauge fields (while again dropping Lorentz indices and also restoring the different gauge couplings here for more clarity, ie, $A_a\to g_a\tilde A_a$, \etc.) as
\begin{equation}
g_L(T_{3L}W_{3L}+T_{8L}W_{8L})+(L\to R)= g_LT_{3L}\tilde A_L+ g_RT_{3R}\tilde A_{3R}-g_L \frac{Q_L}{2} \tilde B_L -g_R\frac{Q_R}{2}\tilde B_R\,,
\end{equation} 
and we now recall that the vev $v_3$ has $Q_L=-Q_R$ from Table~\ref{vevtab}. For further purposes of demonstration, let us concentrate on the second pair of terms and imagine in this example 
that we live in a model realization where this vev, $v_3$, is significantly greater than all of the other large vevs so that we can neglect their effects. Then we see that one combination of the 
$\tilde B_{L,R}$, \ie, $B_1=c\tilde B_L-s\tilde B_R$, will get a mass 
whereas the second, $B_2=c\tilde B_R+s\tilde B_L$, where $(s,c)= (g_R,g_L)/\sqrt {g_L^2+g_R^2}$, remains massless. Re-writing the covariant derivative terms, we see that these two mass 
eigenstates will couple as (recalling that $\kappa=g_R/g_L \simeq 1$) 
\begin{equation}
\frac{g_L}{\sqrt{1+\kappa^2}}~ \Big[ \frac{Q_L}{2}-\kappa^2 \frac{Q_R}{2}\Big] ~B_1+\frac{g_L\kappa}{\sqrt{1+\kappa^2}}~\Big[ \frac{Q_L}{2}+\frac{Q_R}{2}\Big]~B_2\,,
\end{equation} 
where, restoring all the constants, we can now identify the previously mentioned field, $\tilde B=B_L-B_R \sim B_1$, and the massless $B_2$ field will subsequently become 
(part of) the usual LRM/SM $B-L$/hypercharge gauge boson after further symmetry breaking and mixing. 

It thus appears attractive to examine the matrix $M^2_{HGB-4}$ in some more detail when certain vevs, or sets of them, become large relative to the others forcing certain symmetries to be 
broken while others respected. Of course, we cannot treat any given scenario too seriously but they can provide us with guideposts for the many scenarios that one might imagine.  As was noted 
earlier, given the set of vevs in the Table~\ref{vevtab} above it is impossible to break $SU(3)'$ without also breaking $SU(3)_R$ since the set of vevs  $u_{2-4}$ 
all carry $SU(3)_R$ quantum numbers; the reverse, however, is not true as the vevs $v_{3,5}$ do {\it not} carry $SU(3)'$ quantum numbers. Let us now turn to a specific example of this approach 
which presents a reasonable setup, which we will call Scenario I, generalizing on the discussion above. We have already seen 
that a (very) large $v_3$ will break $3_L3_R\to2_L2_R1_L1_R\to 2_L2_R1_{L+R}$; we also simultaneously observe that if $u=u_2=u_4$ is also very large it will act somewhat similarly 
while simultaneously breaking $SU(3)'$ but will also leave $SU(2)_R$ intact. This being the case, let us consider the limit where $v_3^2,u^2 >> v_5^2,u_3^2$ (but still neglecting the 
$SU(2)_L$-breaking vevs) as a working Scenario; then one finds 
(in this approximation), in addition to the massless field described above, two (very) heavy fields: $\tilde B$, a.k.a, $Z_1$, corresponding to the eigenvector $\sim (0,1,-1,0)^T$ as we might 
have expected, with $M^2_{\tilde Z_1}=g^2(2v_3^2+u^2)/4$,  and also $Z_2 \sim (0,1,1,2)^T$, with $M^2_{Z_2}=3g^2u^2/4$. 
Finally, there is also a somewhat lighter field (though being still quite heavy in comparison to the 
electroweak scale), $Z_3\sim (3,1,1,-1)^T$, with $M^2_{Z_3}=g^2(v_5^2+u_3^2)/3$. Note that none of the $Z_i$ appear very much like either $Z_R,$ or $A'$ due to the rather large amount 
of mixing that one finds in this Scenario I. This is, perhaps, not overly surprising as such large mixings between $Z_R$ and the $SU(2)_I$'s heavy partner of the DP, the $A'$ analog in that 
analysis there termed $Z_I$, was also found in to occur in earlier work\cite{Rizzo:2023qbj}.

To demonstrate the efficacy/transparency of taking the $g_L=g_R=g'$ limit in obtaining the results above, we note that if we had {\it not} assumed the equality of the gauge couplings, then the 
$4\times 4$ eigenvector for the massless HGB field would instead be $\sim (1,-g_R/g_L,-1,g_R/g')^T$ 
and the eigenvector for the gauge field $\tilde B$ would now be $\sim (0,1,\gamma,\delta)^T$ where we have defined 
\begin{equation}
\gamma=-\frac{g_R}{g_L}+\frac{g_R}{g'}\delta, ~~~ \delta=\frac{g'(\lambda-c)}{g_Lg'^2u^2-g_Lg_R^2v_3^2},~~~ c=(g_L^2+g_R^2)v_3^2+g_L^2u^2,~~~d=(g_R^2+2g'^2)u^2\,,
\end{equation} 
and 
\begin{equation}
\lambda=\frac{1}{2}(c+d)+\frac{1}{2}\big[(c-d)^2+T\big]^{1/2}\,,
\end{equation} 
with $T=(g_L^2-g_R^2)(g'^2u^2-g_R^2v_3^2)$ and the corresponding mass-squared eigenvalue would instead have been $M^2_{\tilde Z_1}=\lambda/4$, none of which is very transparent or 
overly useful other than observing that the results are $g'$-independent. We clearly see, however, that as $g_R\to g_L$, \ie, $\kappa \to 1$, one finds that both $T,\delta \to 0$ so that 
$\lambda \to c$ and $\gamma \to -1$ and we staightforwardly recover the previously obtained simplified result.

Due to their interplay in the phenomenology, it is interesting to consider the corresponding set of masses for the NHGB in this same Scenario I under the $g_L=g_R=g'$ simplifying 
assumption employing the set of expressions above and in which limit mixing between these states can be approximately ignored. Here we see in this approximation that $W_R$ is very 
likely to be the lightest amongst these states with $M_{W_R}^2=g^2v_5^2/4$, while for $U_{L,R}$ one finds $M_{U_{L,R}}^2=g^2(u^2+v_3^2+v_5^2)/4, g^2(u^2+v_3^2)/4$, respectively, and 
for $V_{L,R}$, 
$M_{V_{L,R}}^2=g^2(u^2+v_3^2+v_5^2)/4,g^2(u^2+u_3^2+v_3^2+v_5^2)/4$, respectively.  We also see that for $O(1)$ Yukawa couplings, the various new fermions will have masses interspersed 
in this same mass range.  Also we note the equality of the $U_L$ and $V_L$ masses due to the (at this point) unbroken $SU(2)_L$ 
gauge symmetry. Furthermore, before turning on any of the $w_i$ vevs, one finds that $M_{W',V'}^2=g^2(2u^2+u_3^2)/4$ 
and, obviously, the contributions for the $w_i$ will only make these states heavier. Of course, the $U'$ mass is set only by the the values of the $w_i$ themselves.

\begin{figure}[htbp]
\centerline{\includegraphics[width=5.0in,angle=0]{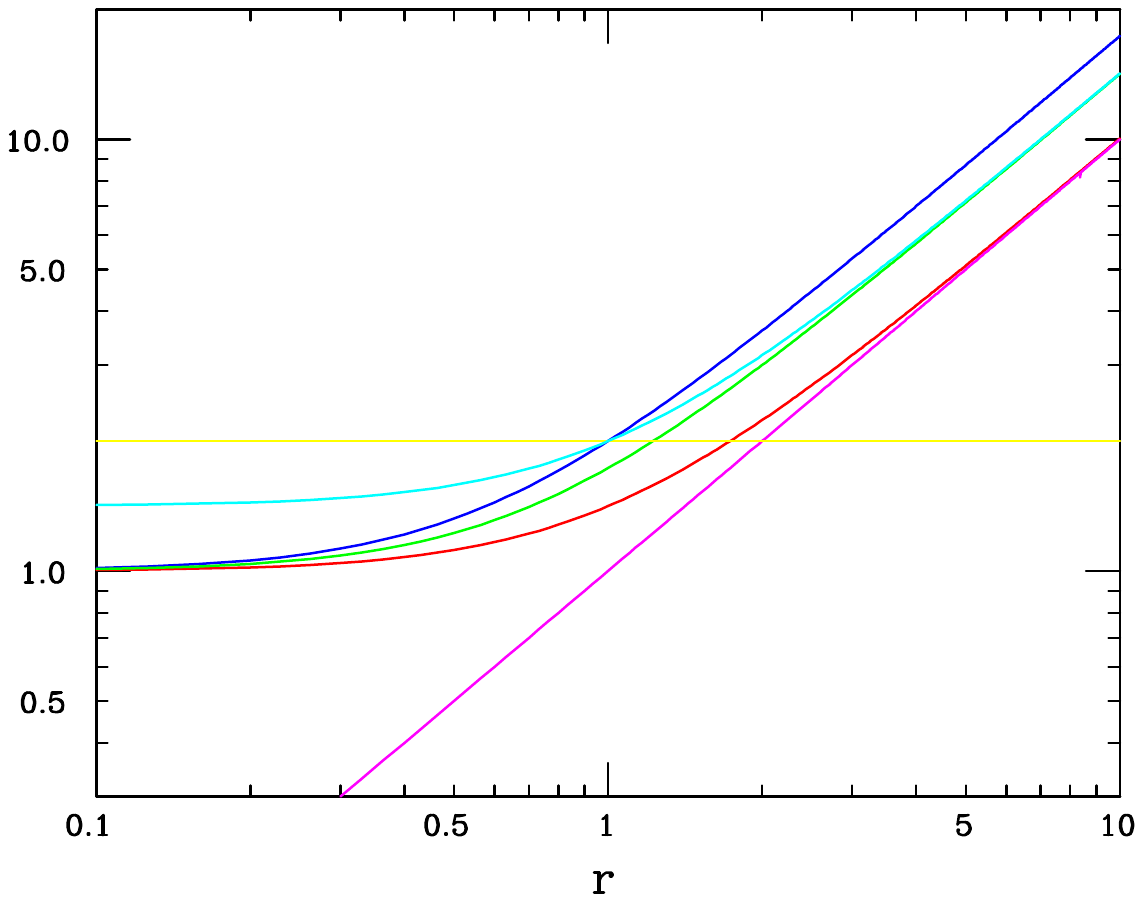}}
\vspace*{-1.3cm}
\caption{Sample mass spectrum of the various NHGB and HGB for Scenario II in units of $M_0=gu/2$ as a function of the vev ratio $r=v/u$, as discussed in the text. For values of $r$ slightly 
above unity, the curves, from top to bottom, correspond to the masses of the $Z_3$ (blue), $V_R$ (cyan), $U_L=V_L$ (green), $Z_1$ (yellow), $U_R=Z_2$ (red), and $W_R$ (magenta), 
respectively.}
\label{spec}
\end{figure}

By way of contrast, we now consider a second Scenario (II) where we imagine a common scale for the large vevs in $H_1$, \ie, $v_3=v_5=v$, which do not break $SU(3)'$, and a similar common 
scale for the large vevs in $H_{2,3}$ that do, \ie, $u_{2-4}=u$, but, to be general, with no specific ordering between $v$ and $u$. In such a case, we find instead the following mass eigenstates, 
still assuming a common gauge coupling for simplicity: $Z_1 \sim (1,-1,-1,-3)^T$ with a mass $M^2_{Z_1}=g^2u^2$, $Z_2 \sim (1,0,1,0)^T$ with mass $M^2_{Z_2}=g^2(v^2+u^2)/4$, and 
$Z_3 \sim (1,2,-1,0)^T$ with $M^2_{Z_3}=g^2(3v^2+u^2)/4$; note that, necessarily, $M^2_{Z_3}>M^2_{Z_2}$ while the mass ordering of $Z_1$ and $Z_3$ depends on whether $u$ or $v$ 
is the larger. Again, we see that, in this case, all three of the very heavy states are quite highly 
mixed with none of them appearing to be much like $Z_R,A'$ or even $\tilde B$ and this result is also observed to be insensitive to any further assumptions with respect to possible 
hierarchies between $v^2$ and $u^2$. In the corresponding NHGB sector, ignoring mixing effects, the $W_R$ again appears as the lightest of these new gauge boson states 
with $M_{W_R}^2=g^2v^2/4$, while for $U_{L,R}$ one finds $M_{U_{L,R}}^2=g^2(u^2+2v^2)/4, g^2(u^2+v^2)/4$, respectively, and now for $V_{L,R}$, one obtains instead 
$M_{V_{L,R}}^2=g^2(u^2+2v^2)/4,g^2(2u^2+2v^2)/4$, respectively.  We again note the equality of the $V_L$ and $U_L$ masses in the limit that $SU(2)_L$ remains unbroken and that the $W_R$ 
tends to be the lightest state. Furthermore, before turning on any of the $w_i$ vevs, in this case one finds that $M_{W',V'}^2=3g^2u^2/4$ with 
the contributions for the $w_i$ further increasing these masses while also providing one for the $U'$. From these considerations (although only based on our experience with this pair of specific 
breaking scenarios) it is clear that the HGB and NHGB will always have overlapping yet hierarchal mass spectra. Though quantitatively different from Scenario I, Scenario II possesses essentially the 
same qualitative features although, apart from the overall scale, the mass spectrum in this case depends upon a single parameter, the vev ratio $r=v/u$, as can be seen in Fig.~\ref{spec}. Of 
course, even in this simplified case, we find that mass relations can be somewhat complex due to this single parameter sensitivity especially when $r\sim 1$ although this situation does simplify 
if either the $r<<1$ or $r>>1$ limits are realized, resulting in some further degeneracy among these new gauge boson states.

Of course, many other orderings of the large scale vevs are possible and will correspond to differing symmetry breaking patterns.

In either of the Scenarios above, and more generally, once the electroweak scale vevs, $v_{1,2,4},u_1$, are turned on, the two, non-DP, massless states ${\cal H}$ and $A_L$  of the $5\times 5$ 
mass squared matrix will mix to form the familiar 
SM fields, $\gamma$ and $Z_{SM}$, as usual and we see that the amount of mass mixing between the $Z$ and the other more massive $Z_i$, which generically live at the few TeV 
scale and above. will always be suppressed by mixing angles which are or order $\sim M_{Z_{SM}}^2/M_{Z_i}^2 \sim 10^{-4}$, consistent with any requirements from the 
(tree-level) electroweak constraints\cite{Workman:2022ynf}. The size of the different $Z_{SM}-Z_i$ mixings will also be somewhat sensitive to the relative magnitudes of the set of  
$v_{1,2,4},u_1$ vevs. From the structure of the $5\times 5$ mass-squared matrix above we see that these same vevs will also induce very small corrections to the 
$Z_i$ masses and the mixings among these HGB beyond those which we've already encountered but which we can safely ignore numerically amongst just these heavy states. Also, 
as noted, once the $SU(2)_L$ gauge symmetry is broken, the degeneracy among some of the NHGB states, \eg, $U_L,V_L$, will be lifted although this mass splitting will remain relatively small. 

Apart from these specific spectrum scenarios, some further intricacies are introduced once the $Q_D\neq 0$ vevs turn on, as they produce the DP mass itself and also induce the usual mass 
mixings between the DP and all of the other more massive HGB states, $Z_{SM}$ and $Z_i$, as was seen in the corresponding case of the NHGB discussed earlier. Since these mass mixings 
are generally of order $M_D^2/M_{Z_{SM,i}}^2$, we see that the dominant one (via the $x_{1,2}$ vevs) is that with $Z_{SM}$ and thus the DP essentially picks up a small $Z_{SM}$-like coupling 
to the SM fermions, which is a familiar and frequently occurring feature of many DP models.  As usual, due to the small magnitudes of these vevs, these newly induced mass mixings will 
generally have very little influence on the masses and couplings of the heavier HGB eigenstates themselves as already discussed. 

One other effect of these $Q_D\neq 0$ vevs, which has also been observed in our earlier work, is to generate a small mixing between the HGB $Z_{SM},Z_i$ states (which all have $Q_D=0$) 
and the electrically neutral NHGB $|Q_D|=1$ states via the hermitian structure $\sim Z_{SM}(V'+V'^\dagger)$, which, \eg, allows them to have new decay paths such as $V'\to Z_{SM}D$, 
somewhat analogous to the $W'\to W_LD$ decay previously discussed. It is to be noted that this particular coupling preserves lepton number as $V'$ has an $L=0$ assignment. Overall, the 
lepton-number violating vevs $x_{2,4}$ have far less direct impact in the HGB sector than in the NHGB one since here such vevs can only appear quadratically (instead of linearly in lowest 
order in the NHGB case) and they are already relatively small, $\lsim 1$ GeV.

As a last point of discussion before ending this Section, we can easily see that there are several reasons why we should already suspect that the current model, as so far described, 
remains incomplete - one might strongly suspect that this at least partially due to the lack of a 
fully unified setup. First, we have already seen the necessity of introducing the additional Higgs scalar field, $\Omega$, beyond those required to generate the corresponding fermion masses, 
to complete the corresponding mass generation for the NHGB gauge fields. This explicitly breaks the apparent $Z_4$ symmetry that this model would otherwise possess. Secondly, we have 
not yet discussed the identity of the DM field itself which, at the very least, must carry $Q_D\neq 0$ while also being an $3_c3_L3_R$ singlet with $Q_{em}=0$, properties not possessed by 
any of the fermion fields or among the many scalars in any of the $H_i$ introduced above. Indeed, as mentioned previously, the CMB constraints on annihilating DM are most easily satisfied when the 
DM is a $p-$wave annihilating complex scalar, $\phi$, which does {\it not} get a vev so as to maintain its stability. Since all of the diagonal members of 
the $\Omega$ Higgs field are seen to obtain vevs to complete the NHGB mass generation process as just discussed, only the off-diagonal field, $\Omega_{13}$, could potentially be a DM 
candidate (albeit with significant fine-tuning), with 
the alternative being to introduce an additional, vev-less, scalar representation with all of the necessary properties. Finally, since the gauge and fermion fields in this setup are completely fixed and 
the Higgs scalar fields necessary to break the gauge symmetries and generate the relevant fermions masses have all been introduced, one might ask whether or not the KM parameter 
$\epsilon$, as was defined above in terms of the field content of the model, is finite or not, \ie, is Eq.(2) satisfied automatically? Certainly this is {\it not} the case given the field content of the 
single fermion generation, $n_g=1$, toy model described above  
as is can be easily seen. While $q,q^c$ and $H_1$ make no contribution here as they all have $Q_D=0$, both $l$ and $l^c$, being color-singlet chiral fermions, will each contribute a factor of 3. 
Correspondingly, the fields $H_{2,3}$, being complex scalars,  will each make an additional contribution of $3/2$ to this sum, assuming that the relevant fields are physical and do not become 
Goldstone bosons. The scalar Higgs field, $\Omega$, on the other hand, also yields a further donation of $3/2$ to the sum (again assuming that all of the contributing scalar fields remain 
physical after SSB) and so the only remaining possible addition to the total, as has been mentioned above, will then arise from the $U'$ and $W'$ NHGB as they carry both $Q_D$ and 
$Q_{em}\neq 0$. To determine their contributions, we follow the work in Ref.~\cite{gauge} which determines that the value of $\eta_i$ for a massive gauge boson with $g=2$ at the tree-level, as 
is the case in renormalizable theories, \ie, $\eta_i=-10$. Hence, these two gauge bosons will then yield a total contribution to the sum of -30. However, we must subtract from this the contribution that 
arises from the relevant eaten Goldstone bosons which we have previously included as part of this sum, an amount equivalent to that obtained from $\Omega$. In any case, it is clear from 
this analysis that the total sum appearing in Eq.(2) is {\it not} identically zero (and is , in fact, found to be equal to -21) implying that $\epsilon$ is not finite in this setup without the presence of 
some (to avoid issues with any gauge anomalies) large number of additional scalar fields. For the general case of an arbitrary number of fermion generations, assuming the same scalar content 
as above, we find that this sum is now given by $6n_g+3-30$ so that for the realistic case of $n_g=3$, we still do not obtain a null result although we are appreciably closer to our goal. Here too, 
a (now somewhat smaller) number of additional scalar fields in the correct representations that are $3_c3_L3_R$ singlets could render $\epsilon$ finite.

\section{Phenomenological Discussion}

The complexities of the current setup are many, mostly due to the signifiant parameter freedom controlling the mass spectrum thus preventing the details of its phenomenology from being 
easily explorable in all generality. However,  as we have seen, specific scenarios with a fixed set of assumptions are clearly much more amenable to such analyses. If anything, this setup is awash in 
new particles at least some of which may be produced at the LHC and/or FCC-hh as well as at future multi-TeV lepton colliders. It predicts eight new NHGB and four new HGB plus a host of 
vector-like fermions, \ie, one new quark, three new charged leptons, four new neutral leptons in addition to a right-handed neutrino for each generation generation as well over a dozen new 
neutral and charged Higgs scalars.  Here we limit our discussion to only a small subset of the new gauge bosons and fermions that we've encountered and which have somewhat complex 
interactions with the conventional SM fields due to their unusual quantum number assignments. While much of this phenomenology will certainly depend upon the 
details of both the mass hierarchy and the mixings among the various new states, in some cases a few model-independent conclusions can be drawn.  Given the complexity of the model, 
in many (if not most) cases, it is also impossible to separate out the consequences of existence of these new gauge bosons from those of the new fermions. This is especially true 
in the limit that the mixings effects induced by the $Q_D \neq 0$ vevs in the various production processes are neglected since many of these new fields connect SM fields to the new ones and 
their relative locations in the mass spectrum are extremely flexible. There is not much new to say in the present context regarding the `conventional' $SU(2)_R$ LRM gauge bosons, 
$W_R, Z_R$, where these particles are close to mass eigenstates except for possible purposes of comparisons,as they has already been the subject of many 
analyses\cite{wpsearch,ATLAS:2022jsi} which, at least semi-quantatively, will still apply to the present situation. Similarly, there is not much new to say about the new isosinglet vector-like quark, 
$h$, as it too is rather `conventional' in that it is by now a well-studied $Q_D=0$ state\cite{Alves:2023ufm}, with the only exceptions being when it participates in the production processes for 
{\it other} new heavy states or it becomes sufficiently heavy to have on-shell, 2-body decays into these new states, \eg, $h\to uW_R^-$. These however, will in many cases will simply mimic some 
of the more familiar SM modes such as $h\to uW_{SM}$. The production of some of the other new states will also mimic familiar production modes that have been well-studied. For example, as 
was noted above, the $eS_1^{(c)}$ final state can be produced either via $\nu-S_1$ and/or $W-U^\dagger$-type mixings from initial state quarks, \ie,  $\bar ud+{\rm h.c}$, similarly to a 
`conventional' heavy neutrino\cite{wpsearch,ATLAS:2022jsi}. Likewise, these same mixings will allow for familiar decays such as, \eg, $S_1\to eW, W\to jj$. The production of 
and signatures for the new PM leptons, $E_i,N_i$, are also already familiar from earlier work, being dominantly pair produced at a hadron/lepton collider via the SM/LRM electroweak gauge 
boson interactions and decaying back into the corresponding SM field plus a DP as was noted above. As noted, these, like $h$, can also act as `intermediaries' or important components of other 
more interesting interactions.

\begin{figure}[htbp]
\centerline{\includegraphics[width=4.8in,angle=0]{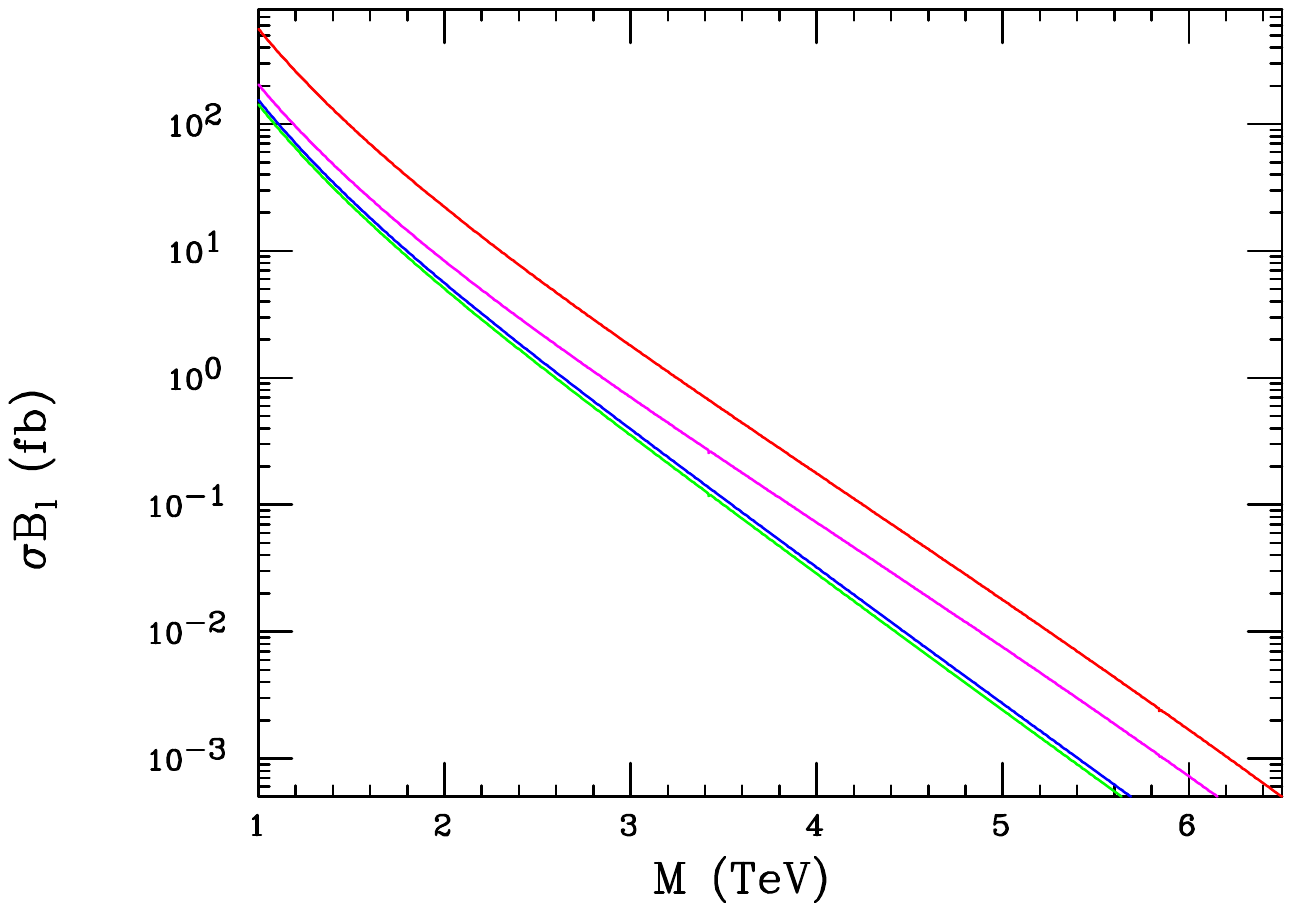}}
\vspace*{-2.0cm}
\centerline{\includegraphics[width=4.8in,angle=0]{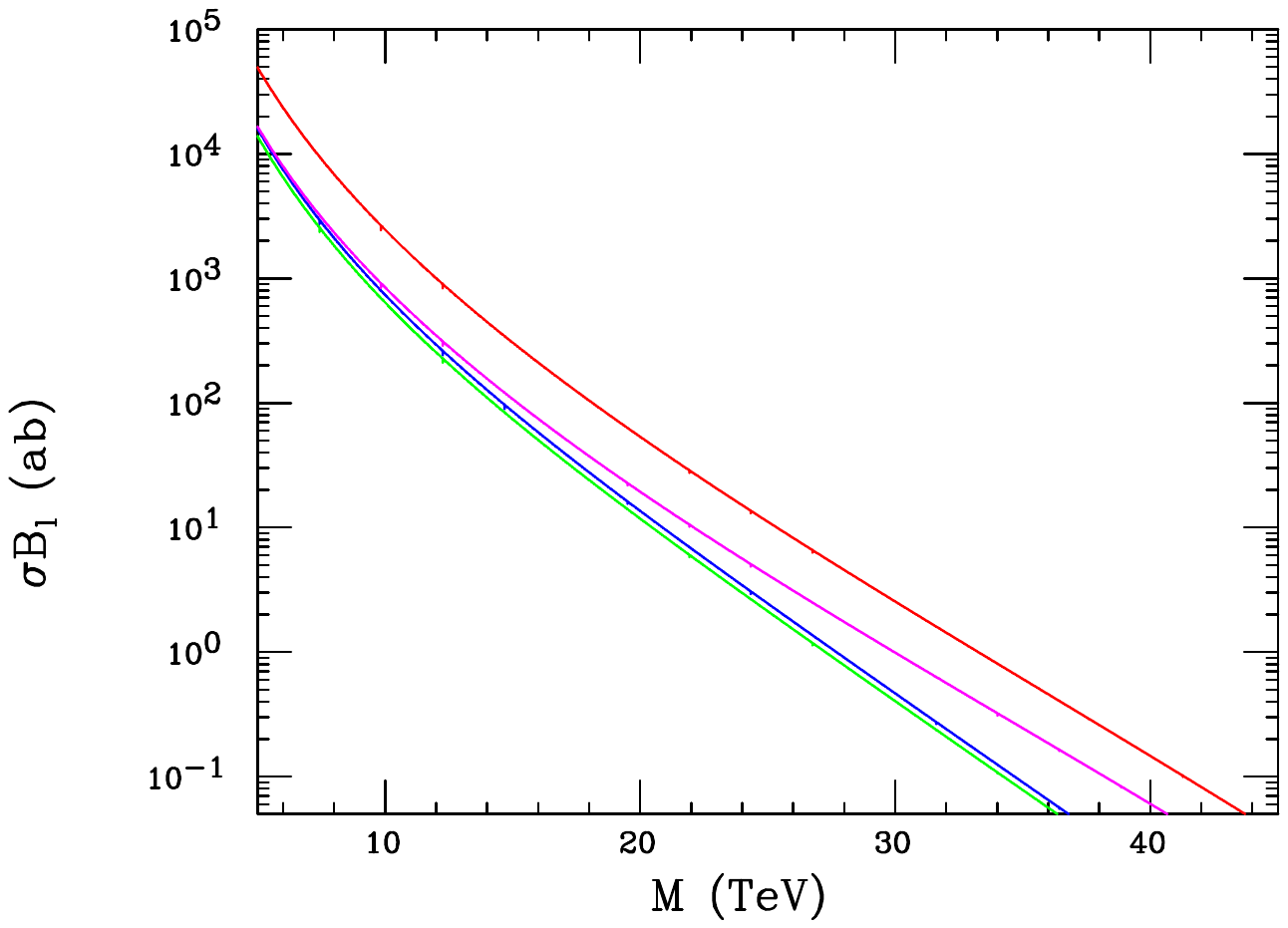}}
\vspace*{-1.3cm}
\caption{(Top) Production cross section times leptonic branching fraction for the three $Z_i$ NHGB in Scenario I in comparison to $Z_{SSM}$ at the 13 TeV LHC as a function of their mass under the 
assumptions as discussed in the text. From top to bottom the curves correspond to $Z_{SSM}$, $Z_3$, $Z_1$ and $Z_2$, respectively. (Bottom) Same as the top panel but now for the 100 TeV 
FCC-hh.}
\label{figm1}
\end{figure}

It is to be noted, given the discussion of the the HGB in the previous Section, that it is not always obvious what is the best basis for making phenomenological predictions involving these states due 
to the rather large amount of mixing involved among the heavier ones (\ie, those outside of $\gamma, D$ and $Z_{SM}$) and one is forced to consider the various specific symmetry breaking 
Scenarios such as I and II above for overall guidance if one wants to make precise predictions. More globally we simply just refer to these three heavy HGB states collectively as $Z_i$ (as above) 
and they will share many common features at the semi-quantitative level outside of certain corners of parameter space. The reason for this is that, fortunately, as we've seen above, the $Z_i$ are 
all sufficiently well mixed implying, \eg, that they all will have some reasonable couplings to the SM quarks, via the $T_{3L,3R}$ and $Y_{L,R}$ generators. This implies that they may be made 
resonantly at a hadron collider and decay into the familiar dilepton final state as is traditionally employed for LHC $Z'$ searches provided that they are kinematically accessible. For example, 
in the case of Scenario I, we see that $Z_1$ is fairly typical in that its (making the standard assumption that only SM final states are kinematically allowed) width to mass ratio is $\simeq 0.017$ 
and it's leptonic branching fraction is $\simeq 0.083$ when $\kappa=1$, values which are not unusual for a new $Z'$\cite{Hewett:1988xc,Leike:1998wr,Langacker:2008yv,Rizzo:2006nw}

Unsurprisingly, the current 13 TeV LHC constraints from ATLAS\cite{ATLAS:2019erb} on the, \eg,  Scenario I $Z_i$'s are not too dissimilar from those on the SSM standard reference model, 
$Z_{SSM}'$, of $\simeq 5.1$ TeV. Following this analysis we obtain a lower bound of roughly $\simeq $4.33 (4.28, 4.72)TeV on the masses 
of $Z_1(Z_2,Z_3)$, respectively, assuming only decays to SM fields in the final state and making the $g_L=g_R=g'$ assumption as employed in the previous Section as can be seen in 
Fig.~\ref{figm1}.  Note that the reaches for the $Z_i$ are somewhat suppressed relative to the usual SSM guidepost at least partially due either the absence or suppression of their couplings to 
up-type quarks. In all cases, however, these reaches may improve by up to $\sim 20-25\%$ 
at the HL-LHC given the significantly greater integrated luminosity and the slightly higher value of $\sqrt s$. For the 100 TeV FCC-hh, employing the same integrated luminosity (30 ab$^{-1}$) 
and search criteria as in Ref.\cite{Helsens:2019bfw,Nemevsek:2023hwx}, we 
find the corresponding reaches of roughly $\simeq $35.2 (34.8,38.9) TeV for the $Z_{1,2,3}$, which are comparable to that of the SSM $Z'$ result of 42 TeV. 
Of course, if the $Z_i$ have additional important non-SM decay modes these search reaches will be somewhat degraded by the lower leptonic branching fraction 
but this possibility also opens up new paths for discovery involving signatures produced by some subset of the other new fields that needed to be introduced in this setup. The HGB in Scenario II 
will semi-quantitatively behave quite similarly.

\begin{figure}[htbp]
\centerline{\includegraphics[width=4.8in,angle=0]{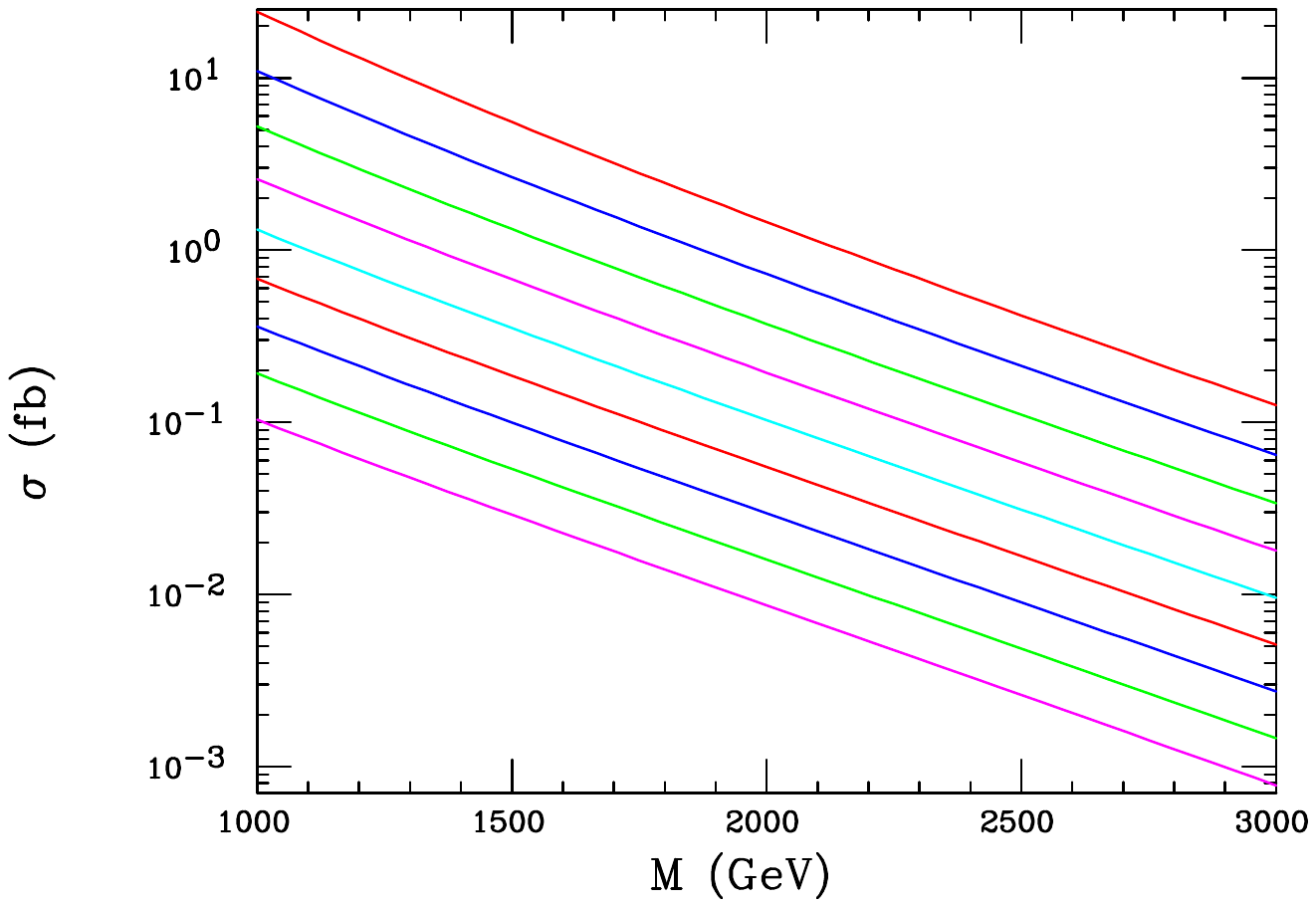}}
\vspace*{-2.0cm}
\centerline{\includegraphics[width=4.8in,angle=0]{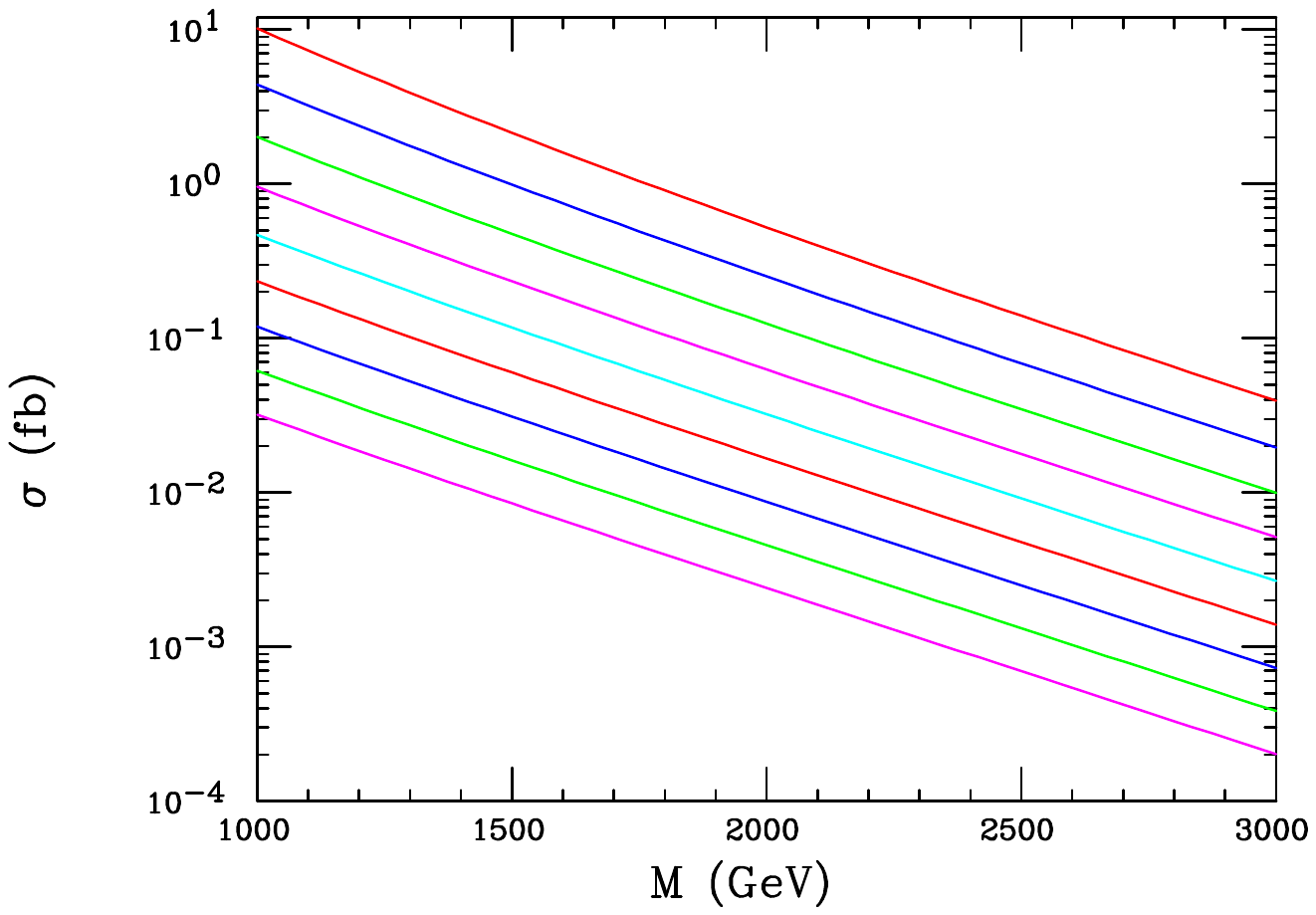}}
\vspace*{-1.3cm}
\caption{(Top) The $gu\to U_{L,R}h+{\rm h.c.}$ associated production cross section as a function of the $U_{L,R}$ mass assuming a SM gauge coupling, $g_R=g_L$ and, from top to bottom, 
that $m_h=1,1.25,...3$ TeV at the 13 TeV LHC. (Bottom) Same as the Top panel,  but now for the $gd\to V_{L,R}h+{\rm h.c.}$ associated production cross section as a function of the $V_{L,R}$ 
mass.}
\label{fig1}
\end{figure}

While these new heavy HGB can be treated reasonably symmetrically, the situation with the NHGB will be somewhat different as generally the mixing amongst most of these states is rather small. 
The NHGB will clearly fall into two main categories: those that are part of the $SU(3)'$ gauge group and those that aren't due to the possible couplings of the SM fields to others that have 
non-zero values of $Q_D$ (in the absence of the $Q_D\neq  0$ vevs). As will be noted, we will omit discussions of the production and signatures of all such states that which are purely of the 
canonical heavy-$W$-like dilepton pair type.

\begin{figure}[htbp]
\centerline{\includegraphics[width=4.8in,angle=0]{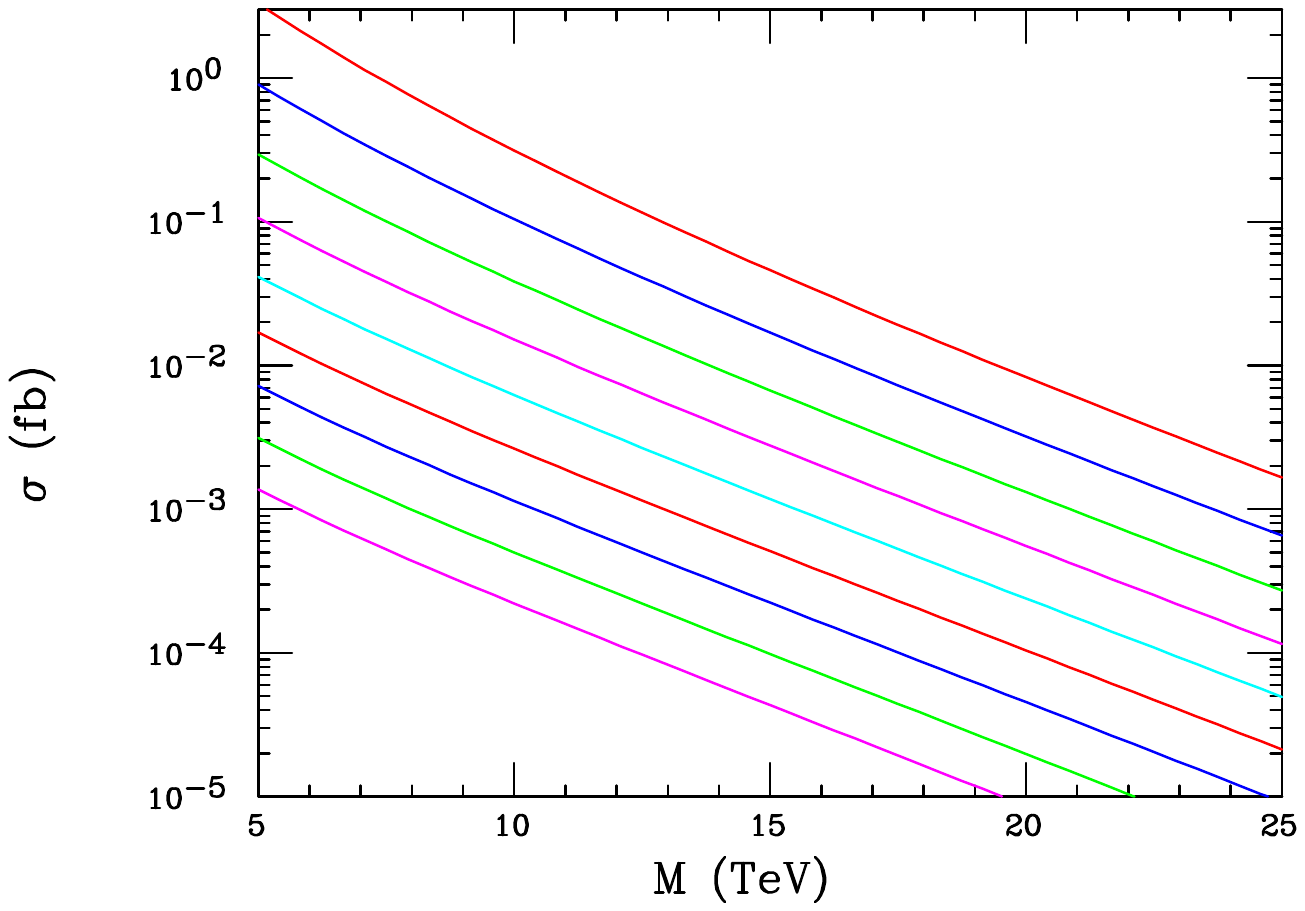}}
\vspace*{-2.0cm}
\centerline{\includegraphics[width=4.8in,angle=0]{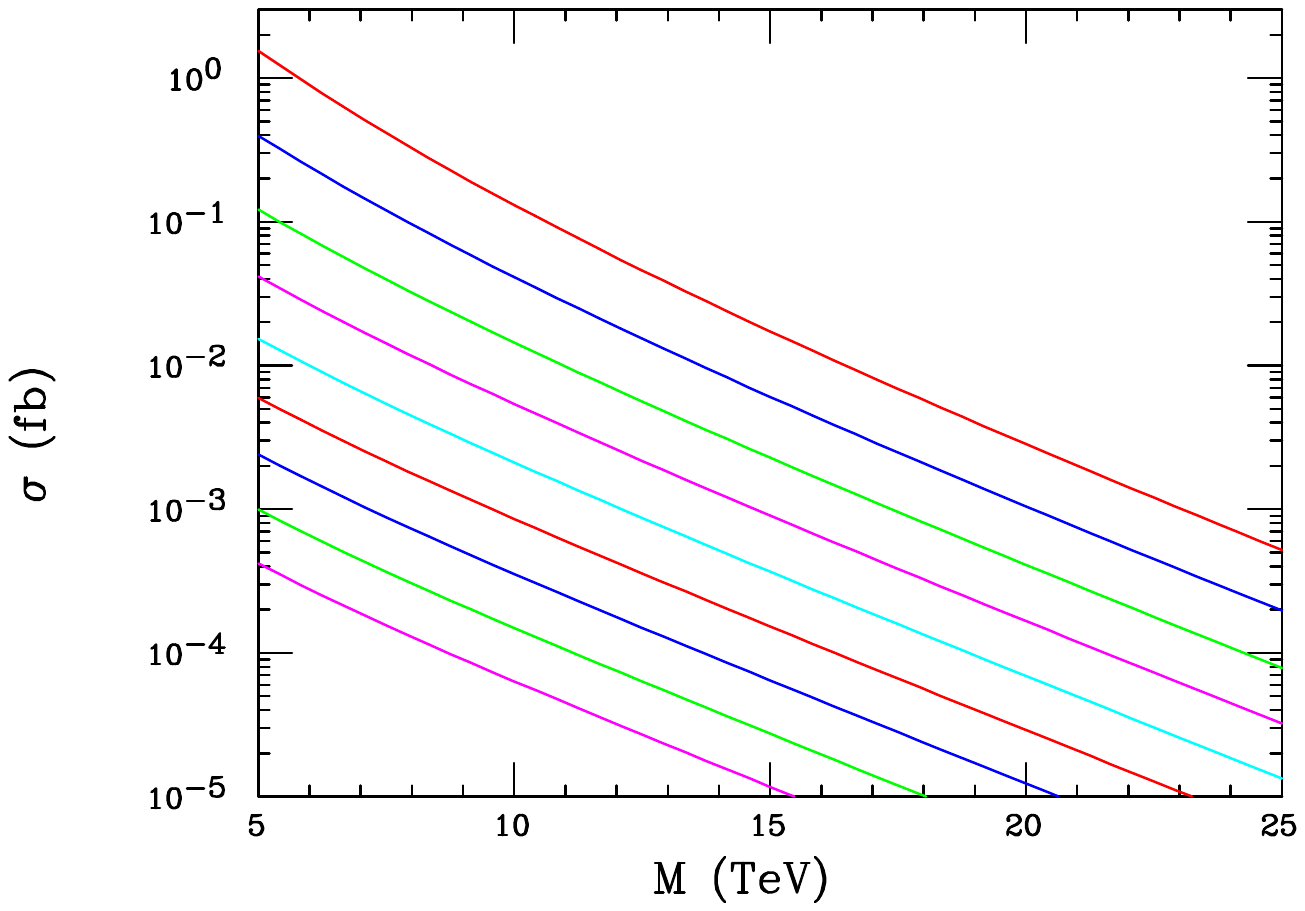}}
\vspace*{-1.3cm}
\caption{Same as the previous Figure but now, from top to bottom, for $m_h=5,7.5,10,...25$ TeV at the 100 TeV FCC-hh.}
\label{fig2}
\end{figure}

Beginning with these NHGB, the states $U_{L,R}$ and $V_{L,R}$ are of immediate interest as, in the absence of mixing with the $SU(3)'$ NHGB, these are the only new ones (apart from $W_R$) 
that will couple directly to the SM quarks and so are more readily made at hadron colliders such as the LHC and FCC-hh. There are two `conventional' mechanisms by which such states may be 
produced in the absence of significant mixing with the other NHGB states (\eg, $W_R$ in the case of $U_R$): associated production together with the heavy $h$ vector-like quark or via 
pair production. Note that in the absence of large mixing these NHGB cannot be produced singly (on their own) as a resonance as can, \eg, the SM $W$ or the $W_R$ in the LRM, since they 
couple to $\bar uh$ and 
$\bar dh$ (and {\it not} to $\bar ud$), respectively. Unlike in previous studies, where the analog of $h$ carried $Q_D\neq 0$, here $h-d$-mixing induced processes such as, \eg, $V_{L,R,'}D$ 
production cannot occur in this setup. In the case of associated production, which is the least model-dependent since it (roughly speaking) only depends upon the particle 
masses, $U_{L,R}$ ($V_{L,R}$) can be made via gluon-quark fusion in the initial state, \ie, $gu(d)\to hU_{L,R}(V_{L,R})+\rm{h.c.}$ and, as might be expected, is only limited by the available 
collider phase space. Figs.~\ref{fig1} and ~\ref{fig2} show the production cross sections for these processes at the 13 TeV LHC and 100 TeV FCC-hh, respectively, for different values of $m_h$, as 
functions of the $U_{L,R}$ and $V_{L,R}$ masses assuming for simplicity that $\kappa=g_R/g_L=1$. The analogous process, $gd \to hD$, does not occur in the present setup since here 
$h$ has $Q_D=0$. 

The corresponding signatures for these production processes will depend, to some extent, upon the ordering of the $h$ mass (and whose decays are well-known) relative to the $U_{L,R},V_{L,R}$ 
NHBG masses as well as on the masses of lepton-like states $S_i, E_i$ and $N_i$ into which the $U$'s and $V$'s might also pair-wise decay to (approximately) conserve $Q_D$. For 
$U_{L(R)}$, the simplest final state will likely be that of $eS_1^{(c)}$ with a clean, very high $p_T$ charged lepton, while for the case of  $V_{L(R)}$, with $S_1^{(c)}\to \nu^{(c)}$, MET will likely 
be the most important part of the signature depending, of course, on how the $S_1^{(c)}$ itself decays.

\begin{figure}[htbp]
\centerline{\includegraphics[width=4.5in,angle=0]{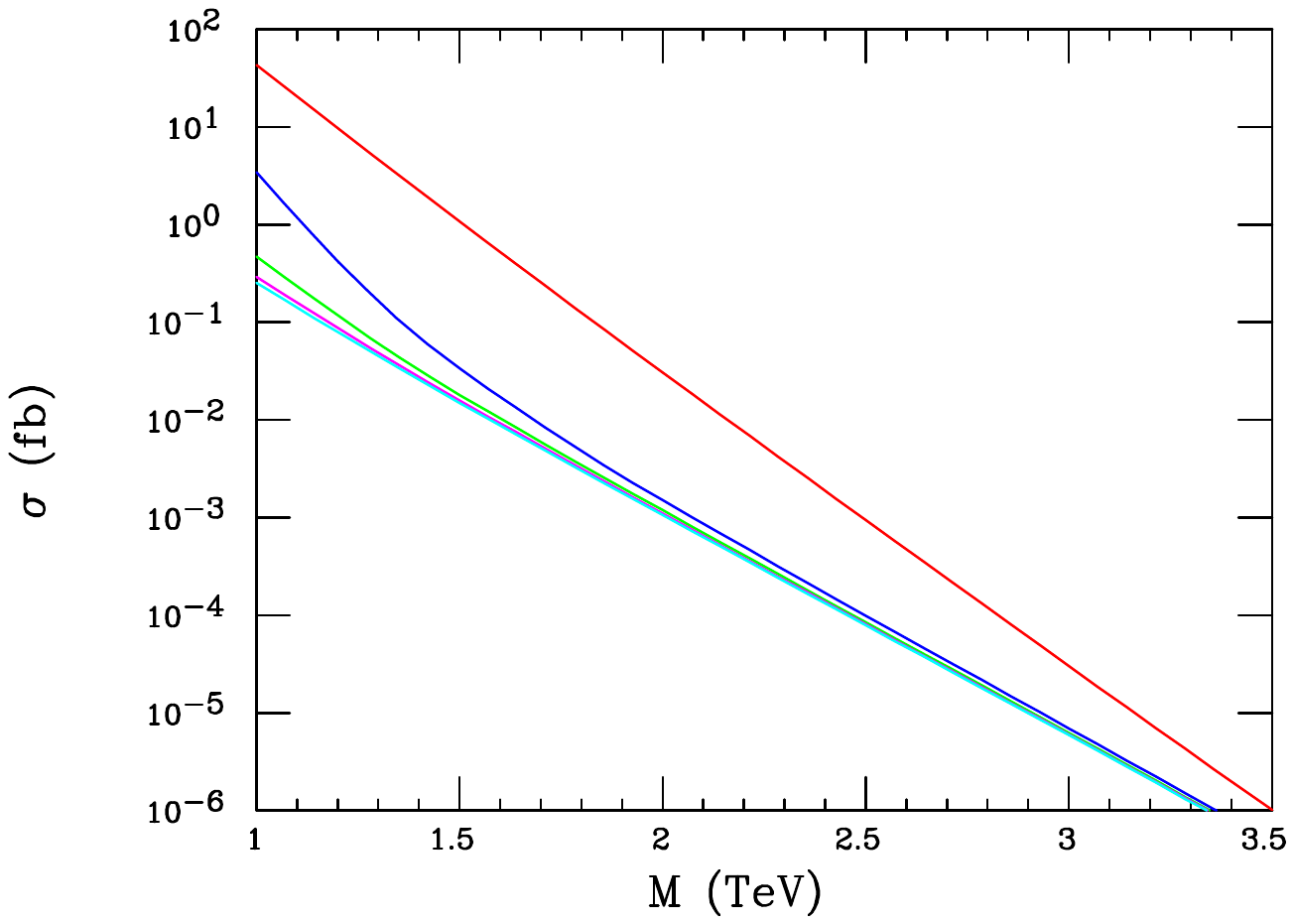}}
\vspace*{-2.0cm}
\centerline{\includegraphics[width=4.5in,angle=0]{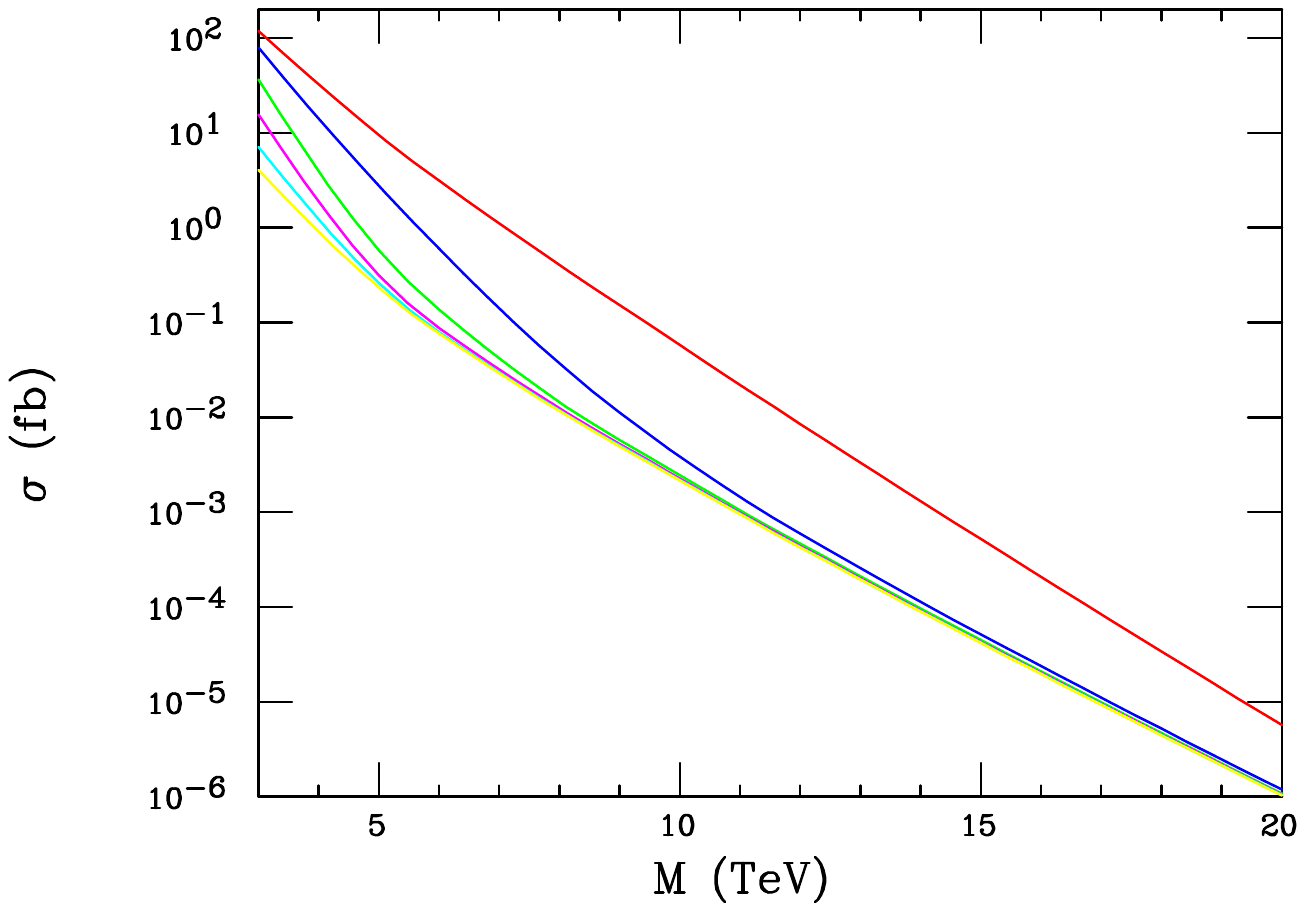}}
\vspace*{-2.0cm}
\centerline{\includegraphics[width=4.5in,angle=0]{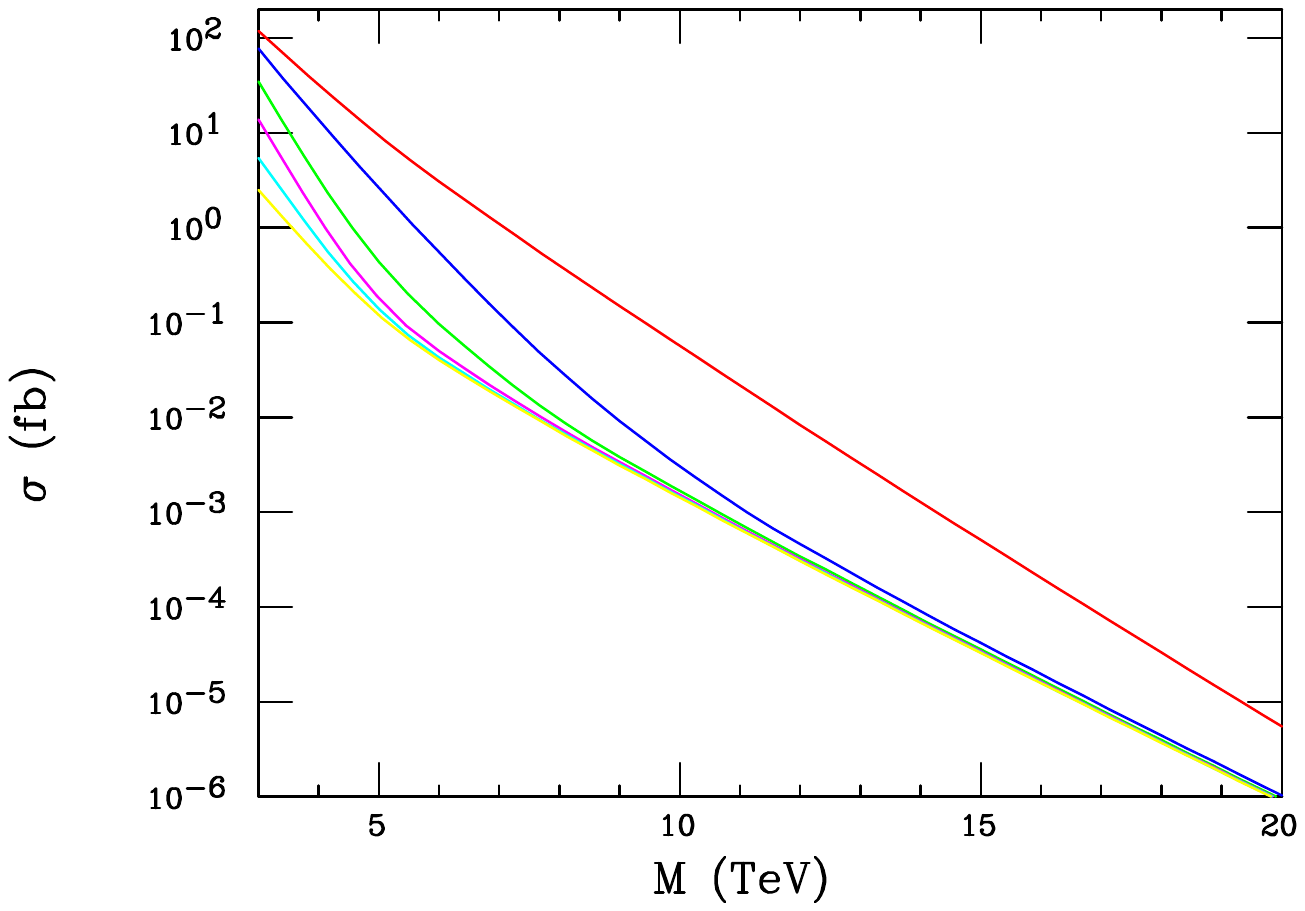}}
\vspace*{-1.3cm}
\caption{$V_{L,R}V_{L,R}^\dagger$ pair production cross section as a function of the mass of $V_{L,R}$, $M$, at (Top) the $\sqrt s=13$ TeV LHC with $m_h=2$ TeV assuming 
$M_{Z'}/M=3,5,7,..$ from top to bottom. Middle (Bottom) Same as the previous panel but now for the $\sqrt s=100$ TeV FCC-hh with $m_h=3(7)$ TeV and assuming the same gauge 
boson mass ratios as above.  $g_R=g_L$ is also assumed in all panels.}
\label{fig0}
\end{figure}
\begin{figure}[htbp]
\centerline{\includegraphics[width=5.0in,angle=0]{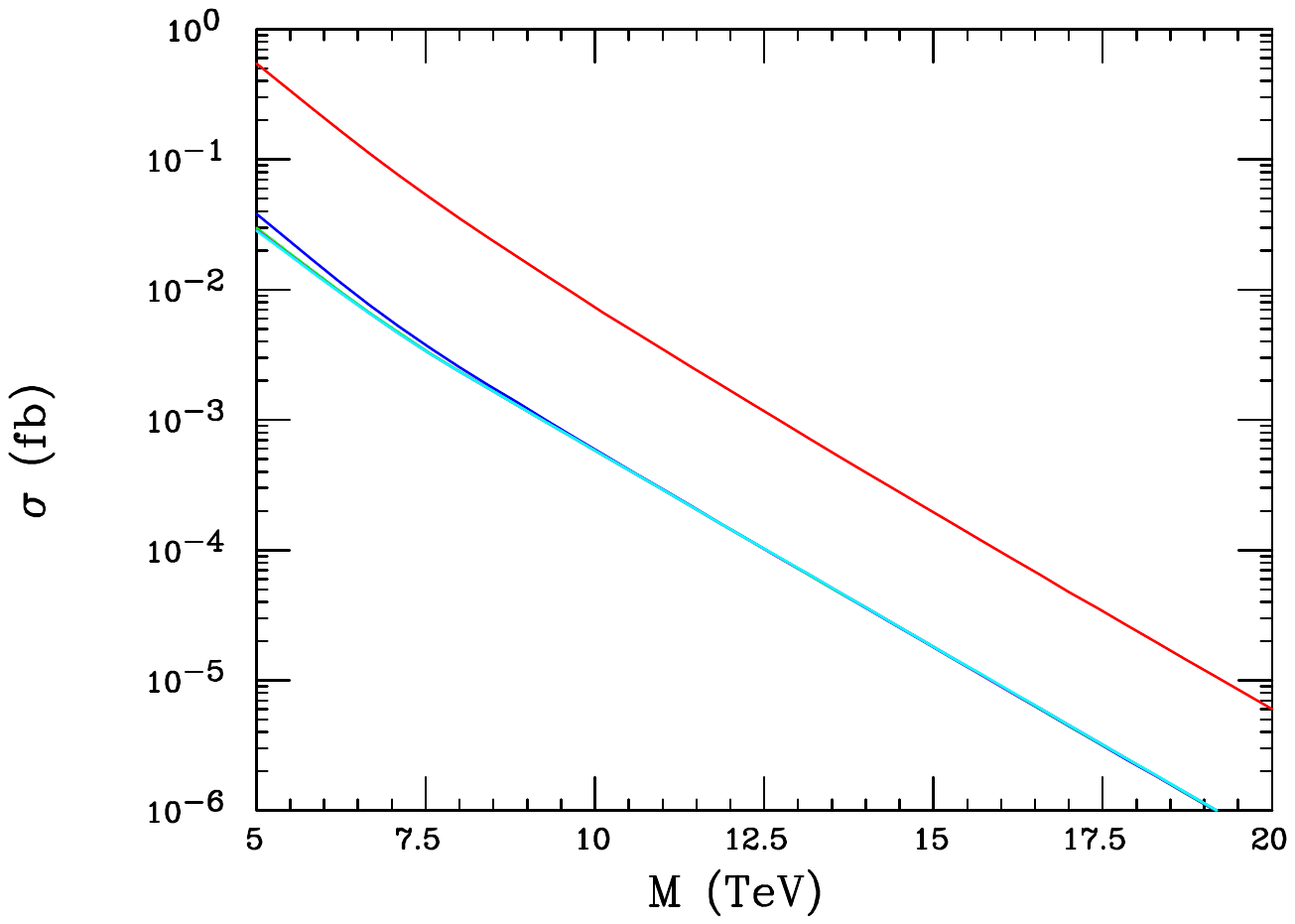}}
\vspace*{-1.3cm}
\caption{Same as in the previous Figure and assuming that $m_h=7$ TeV at the $\sqrt s=100$ TeV FCC-hh but now for $M/M_{Z'}=0.5,1,2,3,4$, respectively, from top to bottom on the left-hand 
side of the panel.}
\label{fig22}
\end{figure}

The second process, $U_{L,R},V_{L,R}$ NHGB pair-production, is the result of $s$-channel exchange (by at least some) of the set of HGBs as well as the $t-$($u-$)channel exchange of the 
$h$ quark, required to maintain unitarity, and is much more highly dependent upon the details of the model than is associated production. Of course in the original weak basis, the only 
contributing $s-$channel exchanges for $U_{L(R)},V_{L(R)}$ pair production are the corresponding $W_{3L(R)},W_{8L(R)}$, but in the mass eigenstate basis the situation much more complex 
due to the rather non-trivial mixing among the $Z_i$ as described above. For example, for the case of $U_{L(R)}U_{L(R)}^\dagger$ production, the $Z_{SM}$, the three $Z_i$, as well as the 
photon will contribute to this process in the $s$-channel. Specifically, in addition to the masses of the $h$ and $U/V$ which appear in the expression for the associated production cross section, 
these additional $s$-channel exchange contributions will also be sensitive to the masses of, \eg,  the $Z_i$ as well as the mixing angles among the weak eigenstates making any 
model-independent predictions for these cross sections impossible. However, the masses and couplings of the various states may be such that one (or more) of the $Z_i$, which is 
kinematically accessible, so can be resonantly produced at a hadron collider, may decay into on-shell pairs of $U$'s and/or $V$'s and so will provide the dominant mechanism for accessing 
these states, with significant cross sections. In both Scenarios I and II, introduced in the previous Section, it was found that the masses of the NHGB and HGB are generally comparable 
and interspersed with each other. For example, in Scenario I, both $Z_{1,2}$ are sufficiently massive to decay into $W_R^+W_R^-$, with $Z_2$ also being allowed to on-shell decay to 
$V_{L,R}V_{L,R}^\dagger$; other modes may be possible depending upon the specific relative magnitudes amongst the various vevs. Scenario II shows a similar pattern although differing 
in the specifics. 

Away from any resonances, it is more difficult to ascertain pair production rates in general due to the many contributing amplitudes. However, as a simple toy example, we again consider the case 
of $V_{L,R}V_{L,R}^\dagger$ pair production far above the SM $Z$ peak but below the $Z_i$ resonance regions and assuming that only one of the $Z_i$'s is the dominant contributor ($Z'$) 
to the cross section. The $h$ exchange amplitude in the $t-$channel is then `fixed' by the specific assumed value of the $h$ mass. This is somewhat similar to what occurs in the previously 
examined $E_6$-inspired model. Fig.~\ref{fig0} shows some examples of this scenario for various values of $M_{Z'}/M_V$ for 
both the $\sqrt s=13$ TeV LHC and the $\sqrt s=100$ TeV FCC-hh where we see this cross section is not very sensitive to the mass of the $h$ since the $Z'$ contribution is resonant.  
As the $Z'$ mass increases relative to that of the $V_{L,R}$, the cross section is found to decrease until $M_{Z'}/M_V\sim 7$ is reached at which point this resonance effect essentially saturates. 
Even though the $Z'$ is resonant, its mass has become sufficiently large that it ends 
up making only a rather small contribution to the overall production cross section. If the $Z'$ mass is such that it cannot (or can barely) resonantly contribute to $V_{L,R}V_{L,R}^\dagger$ 
pair production, \ie, below $2M_V$, we instead obtain results as typically shown in Fig.~\ref{fig22} for the particular case of the FCC-hh. Here we see that while there is some enhancement exactly 
on/near resonance, once $2M_V$ exceeds $M_{Z'}$ this contribution essentially becomes $M_{Z'}$-independent as might be expected. 
In more realistic scenarios, the situation will be somewhat more complex than presented above for numerous reasons, although we might expect that these toy examples have captured at least 
some of the possibilities.

\begin{figure}[htbp]
\centerline{\includegraphics[width=4.8in,angle=0]{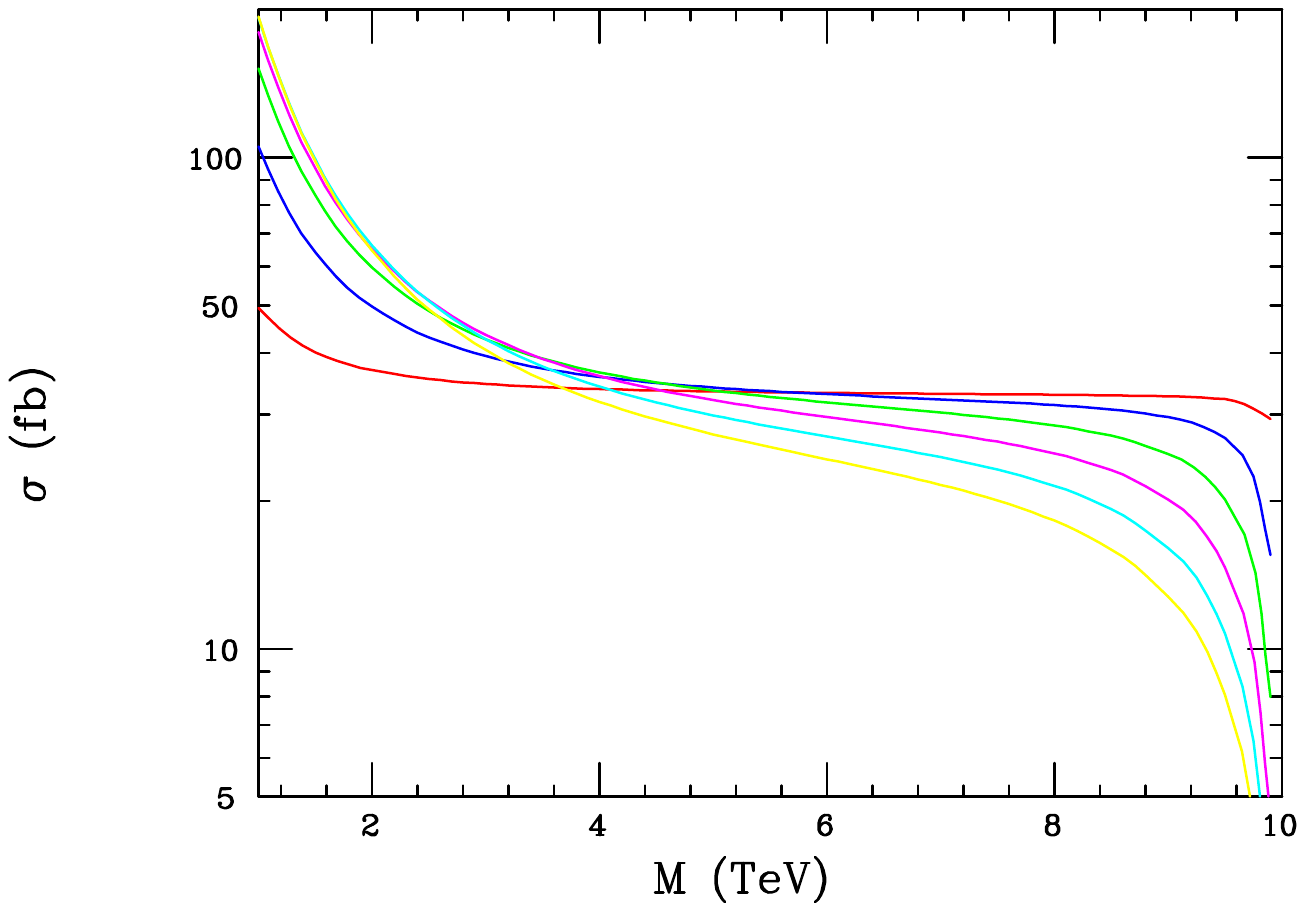}}
\vspace*{-2.0cm}
\centerline{\includegraphics[width=4.8in,angle=0]{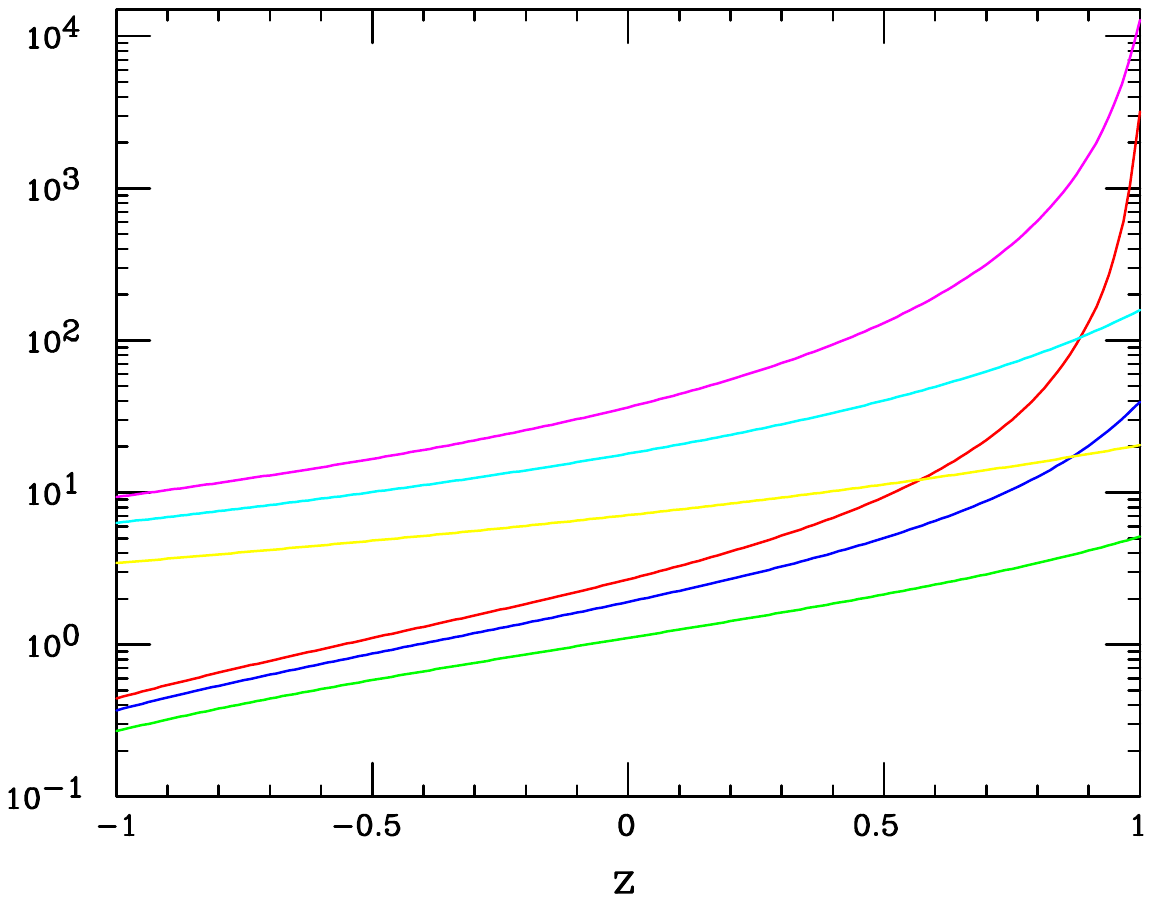}}
\vspace*{-1.3cm}
\caption{ (Top) Production cross section for the $V'D+\rm{h.c.}$ final state at a $\sqrt s=10$ TeV lepton collider as a function of the $V'$ mass $M$ assuming PM lepton masses from 
0.5 to 6.5 TeV from top to bottom on the right-hand side of the panel in steps of 1 TeV. (Bottom) Un-normalized angular distributions, where $z=\cos \theta$, for the same process and 
$\sqrt s$ as in the Top panel 
assuming that, from top to bottom on the left side of the panel, $(M_{V'}, M_E)=(8,1), (8,3), (8,5), (4,1), (4,3)$ and (4,5) TeV, respectively, where $M_E$ is the PM lepton mass. In the Top panel, 
the parameter $\lambda$, as described in the text, is assumed to be equal to unity. }
\label{fig3}
\end{figure}

For the case of the $SU(3)'$ NHGB, in the limit that they remain unmixed with the other NHGB fields, the situation is somewhat different as the SM quarks do not carry any of 
the corresponding $SU(3)'$ quantum numbers so that these gauge bosons are not easily produced at a hadron collider, outside of by pair production.  At lepton colliders, however, other 
processes may are possible,\eg, $e^+e^- \to V'+D$, with $D$ being the DP, via $t-/u-$channel exchange of one or more of the $Q_D \neq 0$, $E_i$ PM fields. Recall from our discussion above 
that the effective $eE_iD$ couplings for 
longitudinal DPs, $\lambda_i$, are both suppressed by the $eE_i$-mixing angles, $\phi_{eE_i} \sim 10^{-4}$, while simultaneously being enhanced by factors of the large mass ratios 
$M_{E_i}/M_D \sim 10^4$ so that we expect $\lambda_i=g \phi_{eE_i}M_{E_i}/M_D \sim O(1)$. This, together with the IR behavior of the amplitude, leads us to anticipate sizable cross sections 
for this process right up to the kinematic production threshold, \ie, $M_{V'}\lsim  \sqrt s$. Fig.~\ref{fig3} shows a sample set of production cross section results for an assumed 
$\sqrt s=10$ TeV lepton collider under the further assumption that a single PM lepton, $E$, exchange is dominant showing that ($i$) the result is not overly sensitive to the the specific PM $E$ 
mass and ($ii$) as might be expected, the cross sections is highly peaked along the beam direction. Once produced, the $V'$ will likely dominantly decay in a not very boosted fashion 
back into the $eE$ final state, assuming it is kinematically allowed, \ie, that $M_{V'}>M_E$, producing a final state with wide-angle, opposite sign leptons. This process can still happen in a 
3-body mode when $E$ is the more massive, $V'\to eE^*\to e^+e^-D$, but in either case the final state will still appear as 2 opposite sign leptons plus missing energy/momentum. 

As noted the $V'+D$ production process cannot occur at a hadron collider via $h$ exchange in the $t/u$-channel but may occur via, \eg, $Z_{SM,i}$ mixing with $V'$ induced by the 
$Q_D\neq 0$ vevs as was mentioned earlier. However, since the relevant mixing angle is likely to be of order $\sim M_D^2/M_{V'}^2$, \ie, quadratic in the small mass ratio unlike in the case 
of fermions or the NHGB, one finds that the rate for this process is numerically suppressed and so is not very useful in probing this setup. We note, however, that if relevant $Z_i$ is more massive 
than $V'$ then it is possible for this suppression to be at least partially offset by a resonance enhancement provided the $Z_i$ is within the kinematically accessible mass range of the collider as 
was noted earlier.

\section{Discussion and Conclusions}

As is well-known, the kinetic mixing of the SM photon and the dark photon allows for the possibility that thermal dark matter, lying in the sub-GeV mass range, can reproduce the observed 
relic abundance of dark matter for a reasonable range of model parameters while still satisfying all other known constraints. However, the generation of such KM relies on the existence of a 
new set of particles, here termed portal matter, that will transform non-trivially under both the SM as well as the DP's $U(1)_D$ gauge groups. Given the electroweak constraints, 
such states will most likely consist of heavy vector-like fermions and/or new scalars, some of which must acquire vevs to break $U(1)_D$ as well as any larger gauge structure into which it may 
be embedded. If such particles do indeed exist and generate this KM portal, what are their properties, how can they be discovered and how can we explore their detailed natures? How will they plus 
the dark photon fit into a more UV-complete theoretical framework with the fields of the SM and what other additional structure is necessary to achieve this? In a recent series of papers, we 
have begun to explore these issues following both bottom-up and top-down approaches to model building by employing the guidance provided by some basic frameworks that will naturally contain 
at least some of the necessary ingredients to construct successful scenarios of this kind.  

In the present paper, we have continued to explore these possibilities, motivated by our earlier analyses of both the $E_6$ and Pati-Salam inspired setups. In the past, we have considered UV 
structures in the form of product groups, \ie, $G=G_{SM}\times G_{Dark}$, where $G_{SM}$ was identified as either the conventional SM, $3_c2_L1_Y$, in the case of $E_6$ or, effectively, 
the LRM, $3_c2_L2_R1_{B-L}$, in the corresponding Pati-Salam setup. For either of these possibilities it was assumed that $G_{Dark}=SU(2)_I\times U(1)_{Y_I}$, again inspired by $E_6$.  In 
both of theses cases, there was at most only a partial symmetry directly linking the full SM and Dark Sectors, but these two setups were found to share some general (and obvious) necessary 
common features: the natural occurrence of both new VL fermions and/or charged scalars that can play the role of PM as well the extended gauge group sector that partly connect SM fields to 
the PM ones at the multi-TeV mass scale. In this paper, we have considered a scenario based on anomaly-free Quartification, a quark-lepton symmetric setup previously 
considered in the literature in other contexts, where we now identify $G_{SM}=3_c3_L3_R$, the familiar Trinification group arising from $E_6$, and where also $G_{Dark}=SU(3)'$ so that 
now $G=[SU(3)]^4$, which clearly displays an obvious symmetry between the visible and Dark sectors that was absent from our previously examined scenarios. Although not a true {\it unification} 
in the very traditional sense, since quarks and leptons remain in different representations and the value of $\epsilon$ remains un-calculable with the minimal field content, this brings us 
much closer to goal of realizing a full UV theory than either of our earlier attempts. In this 
setup, $U(1)_D$ is simply a diagonal subgroup of $SU(3)'$ with $Q_D$ being a linear combination of the two diagonal generators, $T_{3,8}'$, that is uniquely determined by the requirement that SM 
fields will all have $Q_D=0$. However, a general, assumption-free, phenomenological analysis of this type of setup is made somewhat challenging due to the significant, but necessary, augmentation 
of the gauge, fermion and Higgs scalar (containing multiple vevs at various scales) sectors beyond those of the SM or LRM pwhich is viewed as being realized at large mass scales, $\gsim 10$ TeV.  
In particular, we are  confronted by a set of eight new, non-hermitian plus three new, hermitian gauge bosons (in addition to the DP) and, per generation and ignoring color degrees of freedom, 
eight additional VL fermions as well as the RH-neutrino. Of these, one is an `ordinary', $Q_D=0$, color triplet, weak isosinglet quark, $h$, another a $Q_D=0$, isosinglet neutral lepton, $S_1$, 
while the remainder are both charged and neutral leptons, all of which carry dark charges, $E_i,N_i,S_2$. Thus, unusually, the PM in this model necessarily consists solely of only color-singlet 
fields\cite{Wojcik:2023ggt}, which for both scalar and fermion result in relatively low discovery reaches for these new states at hadron colliders. While both $h$ and $S_1$ are fairly conventional 
VL fermion states and will decay to SM fields via the usual SM (or LRM) gauge and Higgs bosons, \eg, $h\to dZ$, the PM fermion fields, as we had found in earlier work, will instead dominantly 
decay to a corresponding analog SM field plus the DP, \eg, $E_i\to eD$. Uniquely in the present setup, two of the NHGB, $W',U'$, can {\it also} act as PM since they carry non-zero values of 
both $Q_D$ and $Q_{em}$; this is correlated with the fact that $U(1)_D$ is an abelian subgroup of $SU(3)'$ and that $Q_{em}$ also partially depends upon the same two $SU(3)'$ diagonal 
generators. Since the setup is only partially unified, $\epsilon$ is not finite and calculable if only the minimal particle content is present but can be made so for the case of 3 fermion generations 
with the addition of several complex scalar fields which can be chosen to be  $3_c3_L3_R$ singlets.

As we have seen, this scenario leads to numerous testable predictions, at least at the semi-quantitative level, which are the result of the extended gauge structure, the existence of PM 
fields/new VL fermions, as well as our other model-building requirements, \eg, that the DP's $U(1)_D$ gauge symmetry remains unbroken until the $\sim $ GeV mass scale is reached. As 
we've neglected flavor physics issues in this discussion for simplicity, the primary tests of the current setup will necessarily directly involve these new gauge bosons and fermions and be provided 
by high energy hadron and/or lepton colliders. Since the set of three new heavy HGB, $Z_i$, all couple to various linear combinations of the four group generators 
$T_{3R,8L,8R,3'}$, they will necessarily couple to at least some of the SM quarks and leptons and so, apart from the details discussed above, will generally behave somewhat similarly to the 
many $Z'$ gauge bosons already encountered in the literature. However, if they are sufficiently massive, they may also decay to pairs of both new VL fermions, including PM, as well as some of the 
other gauge bosons, which may be fortuitous as some of these other states can be difficult to produce at colliders by other means with large cross sections. This is especially true for the $SU(3)'$ 
NHGB, $W',U'$ and $V'$, as these states do not directly couple to the SM quarks in the proton (except via mixing with the other NHGB states) and so can only be produced in association with PM 
or in pairs unlike, \eg, the $W_R$ in the LRM. Unfortunately, the mass spectrum and, in particular, the mass ordering of these new gauge bosons (and the VL fermions as well) is highly model 
dependent and can have a significant impact on detailed phenomenological tests of this setup. For example, in the two sample Scenarios, I and II, analyzed above it was observed that the 
$W_R$ was the lightest among the set of new gauge boson states, although this result can easily seen not to be true in all generality but only reflects the particular choices we made with respect 
to the assumed ordering of spectrum of the five multi-TeV vevs.  However, even with this handicap some quite general model signatures were obtainable and analyzed in the previous Sections 
and which will be further discussed elsewhere.

The KM portal remains a very attractive approach to linking the SM and Dark Sectors and providing a window for sub-GeV DM; it is hoped significant experimental evidence for this idea will 
soon be obtained. Further exploration of these types of PM scenarios is clearly necessary.

\section*{Acknowledgements}
The author would like to particularly thank J.L. Hewett for valuable discussions.  This work was supported by the Department of Energy, Contract DE-AC02-76SF00515 
and was performed in part at the Aspen Center for Physics, which is supported by National Science Foundation grant PHY-2210452.



\end{document}